\documentclass[11pt]{article}

\usepackage[T1]{fontenc}
\usepackage[utf8]{inputenc}
\usepackage[final]{acl} 
\usepackage{times}
\usepackage{latexsym}
\usepackage{microtype}
\usepackage{inconsolata}
\usepackage{graphicx}
\usepackage{booktabs}
\usepackage{multirow}
\usepackage{adjustbox}
\usepackage{subcaption}
\usepackage{float}
\usepackage{xcolor}
\usepackage{makecell}
\usepackage{amsmath}
\usepackage{natbib} 
\usepackage{url}    
\usepackage{enumitem}
\usepackage{multicol}
\usepackage{listings}
\setitemize{itemsep=2pt,leftmargin=8pt,topsep=3pt,parsep=0pt,partopsep=3pt}
\setenumerate{itemsep=2pt,leftmargin=12pt,topsep=3pt,parsep=0pt,partopsep=3pt}

\usepackage{titlesec}
\titleformat{\subsubsection}[runin]{\itshape}{\thesubsubsection}{1em}{}
\renewcommand{\thesubsubsection}{\thesubsection.\alph{subsubsection})}

\title{Something's Fishy In The Data Lake: 

A Critical Re-evaluation of Table Union Search Benchmarks}

\author{
  Allaa Boutaleb,
  Bernd Amann,
  Hubert Naacke and Rafael Angarita \\
  \\ 
  \normalsize Sorbonne Université, CNRS, LIP6, F-75005 Paris, France \\
  \texttt{\{firstname.lastname\}@lip6.fr}
}

\date{} 

\begin{document}
\maketitle

\begin{abstract}
Recent table representation learning and data discovery methods tackle table union search (TUS) within data lakes, which involves identifying tables that can be unioned with a given query table to enrich its content. These methods are commonly evaluated using benchmarks that aim to assess semantic understanding in real-world TUS tasks. However, our analysis of prominent TUS benchmarks reveals several limitations that allow simple baselines to perform surprisingly well, often outperforming more sophisticated approaches. This suggests that current benchmark scores are heavily influenced by dataset-specific characteristics and fail to effectively isolate the gains from semantic understanding. To address this, we propose essential criteria for future benchmarks to enable a more realistic and reliable evaluation of progress in semantic table union search.
\end{abstract}

\section{Introduction}

Measurement enables scientific progress. In computer science and machine learning, this requires the creation of efficient benchmarks that provide a stable foundation for evaluation, ensuring that observed performance scores reflect genuine capabilities for real-world tasks. 

Table Union Search (TUS) aims to retrieve tables $C$ from a corpus that are semantically unionable with a query table $Q$, meaning they represent the same information type and permit vertical concatenation (row appending)~\citep{nargesian_tus_2018, fan_table_2023}. As a top-$k$ retrieval task, TUS ranks candidate tables $C$ by a table-level relevance score $R(Q, C)$. This score is typically obtained by aggregating column-level semantic relevance scores $R(C_Q, C_C)$ computed for each column $C_Q$ of the query table $Q$ and each column $C_C$ of the candidate table $C$. The aggregation often involves finding an optimal mapping between the columns of $Q$ and $C$, for instance via maximum bipartite matching~\citep{fan_starmie_2023}. Successful TUS facilitates data integration and dataset enrichment~\citep{khatiwada_santos_2023, castelo_auctus_2021-1}.

Recent research has introduced sophisticated TUS methods with complex representation learning~\citep{fan_starmie_2023, khatiwada_tabsketchfm_2024, chen_hytrel_2023} designed to capture deeper semantics. However, current benchmarks often exhibit excessive schema overlap, limited semantic complexity, and potential ground truth inconsistencies, which raises questions about whether they provide a reliable environment to evaluate advanced TUS capabilities. While state-of-the-art methodologies leverage semantic reasoning to reflect the task specific challenges, observed high performance may be significantly attributed to model adaptation to specific statistical and structural properties inherent within the benchmark datasets. This phenomenon can confound the accurate assessment and potentially underestimate the isolated contribution of improvements specifically targeting semantics-aware TUS.

In this paper, we examine prominent TUS benchmarks \footnote{\label{fn:benchmarks}Preprocessed benchmarks used in our evaluation are available at \url{https://zenodo.org/records/15499092}}, using simple baselines to assess the benchmarks themselves. Our research questions are:
\begin{enumerate}
    \item Do current TUS benchmarks necessitate deep semantic analysis, or can simpler features achieve competitive performance?
    \item How do benchmark properties and ground truth quality impact TUS evaluation?
    \item What constitutes a more realistic and discriminative TUS benchmark?
\end{enumerate}

Our analysis\footnote{Our code is available at: \url{https://github.com/Allaa-boutaleb/fishy-tus}} reveals that simple baseline methods often achieve surprisingly strong performance by leveraging benchmark characteristics rather than demonstrating sophisticated semantic reasoning. Our contributions include:
\begin{itemize}
    \item A systematic analysis identifying limitations in current TUS benchmarks.
    \item Empirical evidence showing simple embedding methods achieve competitive performance.
    \item An investigation of ground truth reliability issues across multiple TUS benchmarks.
    \item Criteria for developing more realistic and discriminative benchmarks.
\end{itemize}

\section{Related Work} 
\label{sec:related_work}
We review existing research on TUS methods and the benchmarks used for their evaluation, with a focus on how underlying assumptions about table unionability have evolved to become increasingly nuanced and complex.

\subsection{Methods and Their  Assumptions}\label{sec:related_tus_methods}

\subsubsection{Foundational Approaches:}
Following early work on schema matching and structural similarity~\citep{sarma2012finding}, \citet{nargesian_tus_2018} formalized TUS by assessing attribute unionability via value overlap, ontology mappings, and natural language embeddings. \citet{bogatu_d3l_2020} incorporated additional features (e.g., value formats, numerical distributions) and proposed a distinct aggregation method based on weighted feature distances. Efficient implementations of these methods rely on Locality Sensitive Hashing (LSH) indices and techniques like LSH Ensemble~\citep{zhu_lsh_2016} for efficient table search.

\subsubsection{Incorporating Column Relationships:}
Beyond considering columns individually, \citet{khatiwada_santos_2023} proposed SANTOS, which evaluates the consistency of inter-column semantic relationships (derived using an existing knowledge base like YAGO~\citep{yago4} or by synthesizing one from the data itself) across tables to improve TUS accuracy.

\subsubsection{Deep Table Representation Learning:}
Recent approaches use deep learning for tabular understanding. Pylon \citep{cong_pylon_2023} and Starmie \citep{fan_starmie_2023} use contrastive learning for contextualized column embeddings. \citet{hu_autotus_2023} propose AutoTUS, employing multi-stage self-supervised learning. TabSketchFM \citep{khatiwada_tabsketchfm_2024} uses data sketches to preserve semantics while enabling scalability. Graph-based approaches like HEARTS \citep{boutaleb2025hearts} leverage HyTrel \citep{chen_hytrel_2023}, representing tables as hypergraphs to preserve structural properties.

\subsection{Benchmarks and their Characteristics} \label{sec:related_benchmarks}

\begin{table*}[h]
\centering
\footnotesize
\renewcommand{\arraystretch}{1.05}
\setlength{\tabcolsep}{4pt}
\begin{tabular}{@{}cl|rrrrr|rrrr|r@{}}
\hline
\multicolumn{2}{c|}{\textbf{Benchmark}} & \multicolumn{5}{c|}{\textbf{Overall Statistics}} & \multicolumn{4}{c|}{\textbf{Column Type (\%)}} & \textbf{Size (MB)} \\
\cline{3-11}
\multicolumn{2}{c|}{} & \textbf{Files} & \textbf{Rows} & \textbf{Cols} & \textbf{Avg Shape} & \textbf{Missing\%} & \textbf{Str} & \textbf{Int} & \textbf{Float} & \textbf{Other} & \\
\hline
\multirow{2}{*}{\textbf{SANTOS}}
& NQ & 500  & 2,736,673 & 5,707 & 5473 × 11 & 9.96 & 65.39 & 17.00 & 11.46 & 6.15 & \textit{$\sim$422} \\
& Q  & 50   & 1,070,085 & 615 & 21402 × 12 & 5.79 & 73.17 & 15.93 & 8.46 & 2.44 & \\
\hline
\multirow{2}{*}{\textbf{TUS$_{\text{Small}}$}}
& NQ & 1,401 & 5,293,327 & 13,196 & 3778 × 9 & 6.77 & 85.43 & 5.93 & 4.77 & 3.86 & \textit{$\sim$1162} \\
& Q  & 125 & 577,900 & 1,610 & 4623 × 13 & 6.86 & 82.05 & 7.08 & 5.84 & 5.03 & \\
\hline
\multirow{2}{*}{\textbf{TUS$_{\text{Large}}$}}
& NQ & 4,944 & 8,416,415 & 53,133 & 1702 × 11 & 12.53 & 90.12 & 5.10 & 3.57 & 1.21 & \textit{$\sim$1459} \\
& Q  & 100 & 213,229 & 1,792 & 2132 × 18 & 14.87 & 90.46 & 3.68 & 4.13 & 1.73 & \\
\hline
\multirow{2}{*}{\textbf{PYLON}}
& NQ & 1,622 & 85,282 & 16,802 & 53 × 10 & 0.00 & 58.74 & 25.36 & 15.90 & 0.00 & \textit{$\sim$22} \\
& Q  & 124 & 11,207 & 880 & 90 × 7 & 0.00 & 75.68 & 22.95 & 1.36 & 0.00 & \\
\hline
\multirow{2}{*}{\textbf{UGEN$_{\text{V1}}$}}
& NQ & 1,000 & 7,609 & 10,315 & 8 × 10 & 5.79 & 91.68 & 3.27 & 4.29 & 0.76 & \textit{$\sim$4} \\
& Q  & 50 & 405 & 546 & 8 × 11 & 5.87 & 90.48 & 4.58 & 4.21 & 0.73 & \\
\hline
\multirow{2}{*}{\textbf{UGEN$_{\text{V2}}$}}
& NQ & 1,000 & 18,738 & 13,360 & 19 × 13 & 8.16 & 82.40 & 11.71 & 5.50 & 0.39 & \textit{$\sim$8} \\
& Q  & 50 & 5,363 & 665 & 107 × 13 & 4.14 & 84.96 & 10.23 & 2.41 & 2.41 & \\
\hline
\multirow{2}{*}{\textbf{LB-OpenData}}
& NQ & 4,832 & 351,067,113 & 89,757 & 72655 × 19 & 3.44 & 52.50 & 22.56 & 22.37 & 2.57 & \textit{$\sim$80834} \\
& Q  & 3,138 & 238,576,481 & 61,815 & 76028 × 20 & 2.90 & 40.60 & 26.28 & 27.60 & 5.53 & \\
\hline
\multirow{2}{*}{\textbf{LB-Webtable}}
& NQ & 29,686 & 1,039,347 & 387,432 & 35 × 13 & 0.01 & 61.07 & 26.28 & 12.64 & 0.01 & \textit{$\sim$170} \\
& Q  & 5,488  & 335,187  & 56,174  & 61 × 10 & 0.00 & 40.43 & 43.06 & 16.51 & 0.01 & \\
\hline
\end{tabular}
\caption{Table Union Search Benchmarks Summary. NQ = Non-query table, Q = Query table.}
\label{tab:benchmark-stats}
\end{table*}

Benchmark creators make design choices at every stage of the construction process that reflect their understanding and assumptions about how and when tables can and should be meaningfully combined. We identify three primary construction paradigms applied for building TUS benchmarks:

\subsubsection{Partitioning-based:} 
\textsc{Tus}$_{\text{Small}}$ and \textsc{Tus}$_{\text{Large}}$ \citep{nargesian_tus_2018}, as well as the \textsc{Santos} benchmark (referring to \textsc{Santos}$_{\text{Small}}$, as \textsc{Santos}$_{\text{Large}}$ is not fully labeled) \citep{khatiwada_santos_2023} partition seed tables horizontally or vertically, labeling tables from the same original seed as unionable with the seed table. This approach likely introduces significant schema and value overlap, potentially favoring methods that detect surface-level similarity rather than deeper semantic alignment.

\subsubsection{Corpus-derived:} The \textsc{Pylon} benchmark \citep{cong_pylon_2023} curates tables from GitTables \citep{hulsebos2023gittables} on specific topics. While this avoids systematic partitioning overlap, the focus on common topics may result in datasets with a general vocabulary that is well-represented in pre-trained models. This can reduce the comparative advantage of specialized table representation learning and data discovery methods.

\subsubsection{LLM-generated:} \textsc{Ugen} \citep{pal2024alt} leverages Large Language Models (LLMs) to generate table pairs, aiming to overcome limitations of previous methods by crafting purposefully challenging scenarios, including hard negatives. However, this strategy introduces the risk of ground truth inconsistency, as LLMs may interpret the criteria for unionability differently across generations, affecting label reliability.

\subsubsection{Hybrid approaches:} \textsc{LakeBench} \citep{deng_lakebench_2024} uses tables from OpenData\footnote{\url{https://data.gov/}} and WebTable corpora\footnote{\url{https://webdatacommons.org/webtables/}} alongside both partitioning-based synthetic queries and real queries sampled from the corpus. However, such hybrid approaches can inherit the limitations of their constituent methods: partitioning still risks high overlap, candidate-based labeling may yield incomplete ground truth, and the large scale of these benchmarks can introduce practical evaluation challenges.

\section{Methodology} \label{sec:methodology}

As TUS methods become increasingly sophisticated, the benchmarks used for their evaluation may contain inherent characteristics that hinder the accurate assessment of progress in semantic understanding. This section outlines our approach to examining prominent TUS benchmarks through analysis of their construction methods and strategic use of simple baselines as diagnostic tools. The goal of advanced TUS methods is to capture deep semantic compatibility between tables, beyond simple lexical or structural similarity. Our investigation first analyzes the various benchmark construction processes to identify potential structural weaknesses, then employs computationally inexpensive baseline methods to reveal how these characteristics enable alternative pathways to high performance, thereby influencing evaluation outcomes. 

\subsection{Analyzing Benchmark Construction} \label{sec:benchmark_analysis_hypotheses}

We examine five prominent families of TUS benchmarks and formulate hypotheses about their potential limitations based on their construction methodologies (Table \ref{tab:benchmark-stats}). We identify three issues stemming from these methodologies: \textbf{(1) excessive overlap}, \textbf{(2) semantic simplicity}, and \textbf{(3) ground truth inconsistencies}, which we detail below:

\subsubsection{Excessive Overlap:}\label{sec:overlap_artifact}

Benchmarks like \textsc{Tus}$_{\text{Small}}$, \textsc{Tus}$_{\text{Large}}$, \textsc{Santos}, and the \textit{synthetic} query portion of the \textsc{LakeBench} derivatives are created by partitioning seed tables horizontally and vertically, with tables derived from the same original seed designated as unionable pairs. We hypothesize that this methodology inherently leads to significant overlap in both schema (column names) and content (data values) between query tables and their ground truth unionable candidates.

To quantify this, we measure overlap using the Szymkiewicz–Simpson coefficient for exact column names ($\mathit{Overlap}_c$, Eq.~\ref{eq:overlap_col}) and for values of a given data type $d$ ($\mathit{Overlap}_v$, Eq.~\ref{eq:overlap_val}) between ground truth pairs.
\begin{align}
    \mathit{Overlap}_c(Q, C) &= \frac{|Cols_Q \cap Cols_C|}{\min(|Cols_Q|, |Cols_C|)}
    \label{eq:overlap_col}
\\
    \mathit{Overlap}_v(Q, C) &= \frac{|V^d_Q \cap V^d_C|}{\min(|V^d_Q|, |V^d_C|)}
    \label{eq:overlap_val}
\end{align}
where $Cols_Q$ and $Cols_C$ denote the sets of column names in the query table $Q$ and candidate table $C$ respectively, and $V^d_Q$, $V^d_C$ represent the sets of unique values of data type $d$ in each table. The coefficient equals 1.0 when one set is fully contained within the other.
\begin{figure}[htbp]
    \centering
    \includegraphics[width=\linewidth]{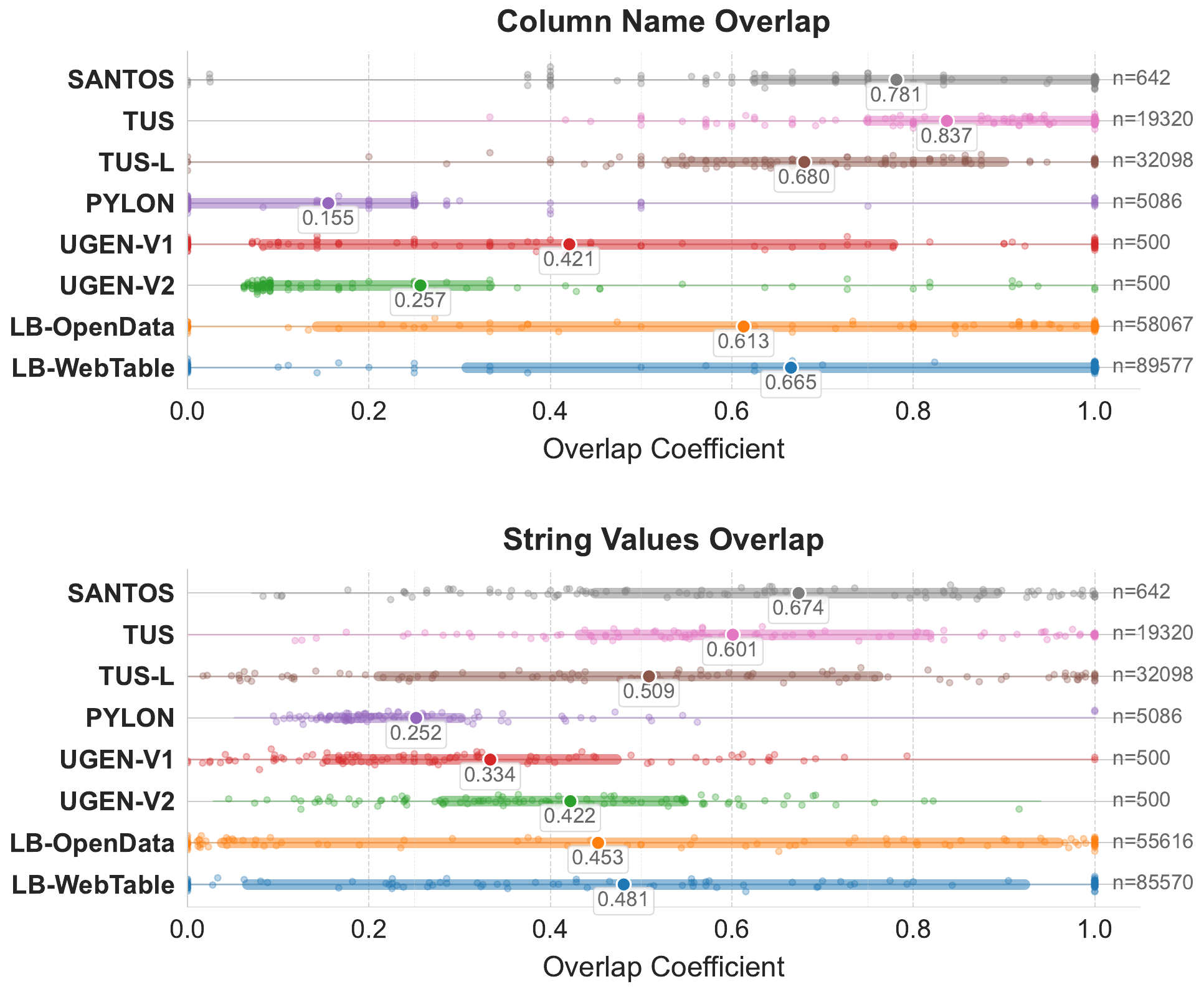} 
    \caption{Distribution of Exact Column Name Overlap (Top) and String Value Overlap (Bottom) Coefficients for Ground Truth Unionable Pairs Across Benchmarks. Colored circles represent mean values; numbers on the right indicate total pairwise relationships considered.}
    \label{fig:benchmark_overlap}
\end{figure}
Figure~\ref{fig:benchmark_overlap} shows the distribution of overlap coefficients, with values $\ge 50\%$ indicating substantial overlap. As expected, partitioning-based benchmarks exhibit high overlap: over 90\% of ground truth pairs share $\ge 50\%$ of exact column names. For value overlap, we focus on string data types, which dominate the benchmarks (Table~\ref{tab:benchmark-stats}). Here too, 45--60\% of query-candidate pairs share $\ge 50\%$ of string tokens. \textsc{LakeBench} derivatives (\textsc{LB-OpenData}, \textsc{LB-Webtable}) show similar trends. Appendix~\ref{appendix:overlap} provides a detailed breakdown by data type. This high surface similarity favors simple lexical methods and also influences advanced models by introducing repeated patterns in serialized inputs (Starmie), data sketches (TabSketchFM), and graph structures (HEARTS). Though designed for deeper semantics, these models are affected by strong benchmark-induced surface signals, making it hard to attribute performance gains purely to nuanced reasoning.

\subsubsection{Semantic Simplicity:}
Benchmarks derived directly from large corpora, such as \textsc{Pylon}~\citep{cong_pylon_2023} using GitTables~\citep{hulsebos2023gittables} or the \textit{real} query portions of \textsc{LakeBench} derivatives using diverse public datasets, avoid the systematic overlap introduced by partitioning. However, we hypothesize that this construction method introduces other limitations since (1) it often focuses on relatively common topics with simpler semantics, reducing the need for specialized domain knowledge, and (2) it generally draws from public data sources likely included in the pre-training corpora of large foundation models.
Evidence from specific benchmarks supports this concern. \textsc{Pylon}'s construction indeed avoids high overlap (Figure~\ref{fig:benchmark_overlap} shows lower overlap than partitioning-based benchmarks). For \textsc{LakeBench}, while the distinction between \textit{real} and \textit{synthetic} queries was unavailable during our analysis\footnote{\url{https://github.com/RLGen/LakeBench/issues/9}}, the significant overall observed overlap suggests that synthetic, partitioning-based queries constitute a large portion of the benchmark. The semantic simplicity evident in \textsc{Pylon}'s topics and the public origins of data in both \textsc{Pylon} and \textsc{LakeBench} could favor general-purpose models like BERT~\citep{devlin-etal-2019-bert} or SBERT~\citep{reimers-2019-sentence-bert}, which have with a high, however unverifiable, probability encountered similar content during pre-training. Consequently, the semantic challenge presented by these benchmarks might be relatively low for models with strong general language understanding -- a contrast to documented LLM struggles with non-public, enterprise-specific data characteristics~\citep{bodensohn2025unveilingchallengesllmsenterprise}, potentially allowing off-the-shelf embedding models to achieve high performance without fine-tuning.

\subsubsection{Noisy Ground Truths:} \label{sec:gt_integrity_artifact} Ensuring accurate and complete ground truth labels is challenging, especially with automated generation or large-scale human labeling efforts as used in LLM-generated benchmarks (\textsc{Ugen}) and large human-labeled ones (\textsc{LakeBench} derivatives). We hypothesize that ground truth in these benchmarks may suffer from reliability issues, including incorrect labels (false positives/negatives) or incompleteness (missed true positives). For \textsc{Ugen},  generating consistent, accurate positive and negative pairs (especially hard negatives) is difficult. LLMs might interpret unionability rules inconsistently across generations, leading to noisy labels. For large-scale human labeling with \textsc{LB-OpenData} and \textsc{LB-Webtable}, the process introduces two risks: \textit{incompleteness}, if the initial retrieval misses true unionable tables; and \textit{incorrectness}, if human judgments vary or contain errors despite validation efforts. Evaluating performance on \textsc{Ugen} and \textsc{LakeBench} derivatives thus requires caution. Scores are affected by label noise or incompleteness; low scores reflect ground truth issues and are therefore not solely attributable to benchmark difficulty, while the maximum achievable recall is capped by unlabeled true positives.

\subsection{Baseline Methods for Benchmark Analysis} \label{sec:baselines_strategy}
Based on the hypothesized benchmark issues identified above, we select some simple baseline methods to test benchmark sensitivity to different information types. While the (1) overlap and (2) general semantics limitations can be directly examined through baseline performance, (3) the ground truth integrity issue requires separate validation of labels, which we address in Section~\ref{sec:ground_truth_issues}. Detailed implementation choices for all baseline methods are in Appendix~\ref{appendix:baseline_impl}.

\subsubsection{Bag-of-Words Vectorizers:} \label{sec:lexical_baselines}
To test whether the \textit{Excessive Overlap} enables methods sensitive to token frequency to perform well on partitioning-based benchmarks, we employ standard lexical vectorizers (\texttt{HashingVectorizer}, \texttt{TfidfVectorizer}, and \texttt{CountVectorizer}) from scikit-learn\footnote{\href{https://scikit-learn.org/stable/api/sklearn.feature_extraction.html}{Scikit-Learn Vectorizers Documentation}}. These generate column embeddings based on sampled string values, with a single table vector obtained via \textit{max pooling} across column vectors. These baselines test whether high performance can be achieved primarily by exploiting surface signals without semantic reasoning.

\subsubsection{Pre-trained Sentence Transformers:} \label{sec:semantic_baselines}
To examine whether the \textit{Semantic Simplicity} allows benchmarks from broad corpora to be effectively processed by pre-trained language models, we use a Sentence-BERT model (\texttt{all-mpnet-base-v2}\footnote{\href{https://huggingface.co/sentence-transformers/all-mpnet-base-v2}{all-mpnet-base-v2 on Hugging Face}}) with three column-to-text serializations: (1) SBERT (V+C): input includes column name and sampled values; (2) SBERT (C): input is only the column name; and (3) SBERT (V): input is only concatenated sampled values. Column embeddings are aggregated using mean pooling to produce a single table vector. These baselines assess whether general semantic embeddings, without task-specific fine-tuning, suffice for high performance on benchmarks with general vocabulary.

\section{Experimental Setup} \label{sec:experimental_setup}

To evaluate our hypotheses about benchmark limitations, we employ both simple baseline methods (Section~\ref{sec:baselines_strategy}) and advanced SOTA methods in a controlled experimental framework. This section details the benchmark datasets used, any necessary preprocessing, the comparative methods, and our standardized evaluation approach.

\subsection{Benchmarks} \label{sec:datasets_preprocessing}

Our analysis uses the benchmarks described in Section~\ref{sec:related_benchmarks}, with post-preprocessing statistics summarized in Table~\ref{tab:benchmark-stats}. Most benchmarks were used as-is, but the large-scale \textsc{LakeBench} derivatives (\textsc{LB-OpenData} and \textsc{LB-Webtable}) required additional preprocessing for feasibility and reproducibility. The original datasets were too large to process directly and included practical issues, such as missing files, as well as characteristics that complicated evaluation, such as many unreferenced tables. We removed ground truth entries pointing to missing files (58 in \textsc{LB-Webtable}), and excluded unreferenced tables from the retrieval corpus (removing $\sim$5{,}300 and $>$2.7M files from \textsc{LB-OpenData} and \textsc{LB-Webtable}, respectively). This latter step was done purely for computational feasibility; as a side effect, it simplifies the benchmark by eliminating tables that would otherwise be false positives if retrieved. We also ensured that each query table was listed as a candidate for itself. These steps substantially reduced corpus size while preserving evaluation integrity. The \textsc{LakeBench} variants considered in our study are those available as of May 20, 2025\footnote{\href{https://github.com/RLGen/LakeBench/commit/df7559d37cec6873c1b96b3ceb08b27ed9efd342}{LakeBench commit df7559d used in our study}}. Future updates to the original repository may modify dataset contents, which yield different evaluation results.

Additionally, for \textsc{LB-OpenData}, we created a smaller variant with tables truncated to 1{,}000 rows, which we use in experiments alongside the original version (Table~\ref{tab:embedding-metrics}). For \textsc{Tus}$_{\text{Small}}$ and \textsc{Tus}$_{\text{Large}}$, we followed prior work~\citep{fan_starmie_2023, hu_autotus_2023}, sampling 125 and 100 queries, respectively. For the other benchmarks, all queries were used.

\subsection{Comparative Methods} \label{sec:methods}

To evaluate our baseline methods (Section~\ref{sec:baselines_strategy}), we compare them against key TUS models previously discussed in Section~\ref{sec:related_tus_methods}, focusing on SOTA methods. For each method, we optimize implementation using publicly available code for fairness:
\begin{itemize}
    \item \textbf{Starmie} \citep{fan_starmie_2023}: We retrained the RoBERTa-based model for 10 epochs on each benchmark using recommended hyperparameters and their ``Pruning'' bipartite matching search strategy for generating rankings, which achieves optimal results according to the original paper.

\item \textbf{HEARTS} \citep{boutaleb2025hearts}: We utilized pre-trained HyTrel embeddings \citep{chen_hytrel_2023} with a contrastively-trained checkpoint. For each benchmark, we adopted the best-performing search strategy from the HEARTS repository: Cluster Search for \textsc{Santos},  \textsc{Pylon},  and \textsc{Ugen} benchmarks, and ANN index search with max pooling for the \textsc{Tus} and \textsc{LakeBench} benchmarks.

\item \textbf{TabSketchFM} \citep{khatiwada_tabsketchfm_2024}: Results for the \textsc{Tus}$_{\text{Small}}$ and \textsc{Santos} were reported directly from the original paper, as the pretrained checkpoint was unavailable at the time of our experiments. 
\end{itemize}

These methods represent significant advancements in table representation learning. AutoTUS \citep{hu_autotus_2023} wasn't included due to code unavailability at the time of writing. We provide further implementation details in Appendix~\ref{appendix:sota_impl}.

\subsection{Evaluation Procedure} \label{sec:eval_procedure}

We use a consistent evaluation procedure for all baseline and SOTA methods to ensure fair comparison. Table vectors are generated per method (Section~\ref{sec:baselines_strategy} for baselines; SOTA-specific procedures otherwise) and L2-normalized for similarity via inner product. For similarity search, baseline methods use the FAISS library~\citep{douze2024faiss} with an exact inner product index (\texttt{IndexFlatIP}); each query ranks all candidate tables by similarity. SOTA methods use FAISS or alternative search strategies (Appendix~\ref{appendix:sota_impl}). Following prior work~\citep{fan_starmie_2023, hu_autotus_2023}, we report Precision@k (P@k) and Recall@k (R@k), averaged across queries. Values of $k$ follow prior works and are shown in results tables (e.g., Table~\ref{tab:embedding-metrics}). We also evaluate computational efficiency via offline (training, vector extraction, indexing) and online (query search) runtimes, with hardware details in Appendix~\ref{appendix:hardware}.

\section{Results and Discussion} \label{sec:results_discussion}

Our empirical evaluation revealed significant patterns across benchmarks that expose fundamental limitations in their ability to measure progress in semantic understanding. Tables~\ref{tab:embedding-metrics} and~\ref{tab:embedding-time} present effectiveness and efficiency metrics respectively.

\begin{table*}[h]
\centering
\small
\renewcommand{\arraystretch}{1.2}
\begin{adjustbox}{width=\textwidth}
\begin{tabular}{@{}l|cc|cc|cc|cc|cc|cc|cc|cc|cc@{}}
\hline
\multirow{2}{*}{\textbf{Method}} & 
\multicolumn{2}{c|}{\textbf{SANTOS}} & 
\multicolumn{2}{c|}{\textbf{TUS}} & 
\multicolumn{2}{c|}{\textbf{TUS\(_{\text{Large}}\)}} & 
\multicolumn{2}{c|}{\textbf{PYLON}} & 
\multicolumn{2}{c|}{\textbf{UGEN\(_{\text{V1}}\)}} & 
\multicolumn{2}{c|}{\textbf{UGEN\(_{\text{V2}}\)}} & 
\multicolumn{2}{c|}{\textbf{\textsc{LB-OpenData\(_{1k}\)}}} & 
\multicolumn{2}{c|}{\textbf{\textsc{LB-OpenData}}} & 
\multicolumn{2}{c@{}}{\textbf{LB-WebTable}} \\

\cline{2-19}
 & P@10 & R@10 & P@60 & R@60 & P@60 & R@60 & P@10 & R@10 & P@10 & R@10 & P@10 & R@10 & P@50 & R@50 & P@50 & R@50 & P@20 & R@20 \\
\hline
IDEAL & 1.00 & 0.75 & 1.00 & 0.34 & 1.00 & 0.23 & 1.00 & 0.24 & 1.00 & 1.00 & 1.00 & 1.00 & 0.39 & 1.00 & 0.39 & 1.00 & 0.81 & 0.95 \\
\hline
\multicolumn{19}{@{}l}{\textit{\textbf{Non-specialized Embedding Methods}}} \\
\hline
HASH & \underline{0.98} & \underline{0.74} & \underline{0.99} & \underline{0.33} & \textbf{0.99} & \textbf{0.23} & 0.64 & 0.15 & 0.59 & 0.59 & 0.43 & 0.43 & 0.21& 0.60& 0.21& 0.60& 0.21& 0.25\\
TFIDF & \textbf{0.99} & \textbf{0.74} & \textbf{1.00} & \textbf{0.34} & \textbf{0.99} & \textbf{0.23} & 0.70 & 0.17 & 0.58 & 0.58 & 0.50 & 0.50 & 0.21& 0.61& 0.21& 0.61& 0.23& 0.27\\
COUNT & \textbf{0.99} & \textbf{0.74} & \textbf{1.00} & \textbf{0.34} & \textbf{0.99} & \textbf{0.23} & 0.68 & 0.17 & 0.58 & 0.58 & 0.50 & 0.50 & 0.21& 0.60& 0.21& 0.60& 0.23& 0.27\\
SBERT (V+C) & \underline{0.98} & \underline{0.74} & \textbf{1.00} & \textbf{0.34} & \textbf{0.99} & \textbf{0.23} & \textbf{0.91} & \textbf{0.22} & \textbf{0.61} & \textbf{0.61} & \textbf{0.68} & \textbf{0.68} & \textbf{0.23} & \textbf{0.66}& \textbf{0.23} & \textbf{0.66} & \textbf{0.26}& \textbf{0.31}\\
SBERT (V) & 0.94 & 0.71 & \textbf{1.00} & \textbf{0.34} & \textbf{0.99} & \textbf{0.23} & 0.84 & 0.20 & 0.58 & 0.58 & 0.58 & 0.58 & 0.22& 0.62& 0.22 & 0.62 & 0.25& 0.29\\
SBERT (C) & \underline{0.98} & \underline{0.74} & \textbf{1.00} & \textbf{0.34} & \underline{0.98} & \underline{0.23} & \underline{0.85} & \underline{0.21} & \underline{0.60} & \underline{0.60} & \underline{0.65} & \underline{0.65} & \underline{0.22}& \underline{0.64}& \underline{0.22} & \underline{0.64} & 0.16& 0.20\\
\hline
\multicolumn{19}{@{}l}{\textit{\textbf{Specialized Table Union Search Methods}}} \\
\hline
Starmie & 0.98 & 0.73 & 0.96 & 0.31 & 0.93 & 0.21 & 0.81 & 0.20 & 0.57 & 0.57 & 0.58 & 0.58 & 0.18& 0.51& \textdaggerdbl & \textdaggerdbl & \underline{0.25}& \underline{0.30}\\
HEARTS & \underline{0.98} & \underline{0.74} & \textbf{1.00} & \textbf{0.34} & \textbf{0.99} & \textbf{0.23} & 0.65 & 0.16 & 0.56 & 0.56 & 0.37 & 0.37 & 0.19& 0.61& 0.19 & 0.60& 0.23& 0.28\\
TabSketchFM & 0.92 & 0.69 & 0.97 & 0.32 & \textasteriskcentered & \textasteriskcentered & \textasteriskcentered & \textasteriskcentered & \textasteriskcentered & \textasteriskcentered & \textasteriskcentered & \textasteriskcentered & \textasteriskcentered & \textasteriskcentered & \textasteriskcentered & \textasteriskcentered & \textasteriskcentered & \textasteriskcentered \\
\hline
\end{tabular}
\end{adjustbox}
\caption{Precision and Recall across benchmarks. Highest values in \textbf{bold}, second highest \underline{underlined}. IDEAL represents the maximum possible P@k and R@k achievable for each benchmark at the specified k. \textbf{\textasteriskcentered} : Results unavailable as checkpoint was not publicly accessible. \textbf{\textdaggerdbl} : Not reported due to excessive computational requirements.}
\label{tab:embedding-metrics}
\end{table*}

\begin{table*}[h]
\centering
\small
\renewcommand{\arraystretch}{1.2}
\begin{adjustbox}{width=\textwidth}
\begin{tabular}{@{}l|cc|cc|cc|cc|cc|cc|cc|cc|cc@{}}
\hline
\multirow{2}{*}{\textbf{Method}} & \multicolumn{2}{c|}{\textbf{SANTOS}} & \multicolumn{2}{c|}{\textbf{TUS}} & \multicolumn{2}{c|}{\textbf{TUS$_{\text{Large}}$}} & \multicolumn{2}{c|}{\textbf{PYLON}} & \multicolumn{2}{c|}{\textbf{UGEN$_{\text{V1}}$}} & \multicolumn{2}{c|}{\textbf{UGEN$_{\text{V2}}$}} & \multicolumn{2}{c|}{\textbf{\textsc{LB-OpenData$_{\text{1k}}$}}} & \multicolumn{2}{c|}{\textbf{\textsc{LB-OpenData}}} & \multicolumn{2}{c@{}}{\textbf{LB-WebTable}} \\
\cline{2-19}
 & Offline & Online & Offline & Online & Offline & Online & Offline & Online & Offline & Online & Offline & Online & Offline & Online & Offline & Online & Offline & Online \\
 \hline
\multicolumn{19}{@{}l}{\textit{\textbf{Non-specialized Embedding Methods}}} \\
\hline
HASH & 0m 15s & 0m 0s & 0m 43s & 0m 1s & 1m 45s & 0m 2s & 0m 19s & 0m 1s & 0m 12s & 0m 0s & 0m 14s & 0m 0s & 7m 56s& 0m 31s& 12m 4s & 0m 22s & 6m 3s & 0m 21s \\
TFIDF/COUNT & 0m 53s & 0m 0s & 1m 45s & 0m 1s & 3m 10s & 0m 2s & 0m 22s & 0m 1s & 0m 9s & 0m 0s & 0m 12s & 0m 0s & 22m 22s& 0m 31s& 37m 14s & 0m 21s & 6m 21s & 0m 22s \\
SBERT & 1m 45s & 0m 0s & 3m 30s & 0m 0s & 9m 21s & 0m 15s & 3m 18s & 0m 0s & 1m 41s & 0m 0s & 2m 20s & 0m 0s & 27m 47s & 0m 4s & 82m 13s & 0m 4s & 30m 45s & 0m 3s \\
\hline
\multicolumn{19}{@{}l}{\textit{\textbf{Specialized Table Union Search Methods}}} \\
\hline
STARMIE & 19m 3s & 1m 2s & 4m 24s & 8m 59s & 14m 43s & 20m 29s & 7m 56s & 3m 27s & 2m 8s & 1m 0s & 2m 45s & 1m 45s & 131m 48s & 1220m 53s & \textbf{--} & \textbf{--} & 48m 11s & 1311m 43s \\
HEARTS & 0m 21s & 0m 34s & 1m 1s & 0m 0s & 3m 10s & 0m 0s & 0m 57s & 0m 36s & 0m 23s & 0m 40s & 0m 30s & 0m 35s & 21m 33s & 0m 3s & 76m 12s & 0m 5s & 29m 28s & 0m 3s \\
\hline
\end{tabular}
\end{adjustbox}
\caption{Computational efficiency across benchmarks. Times are averaged over 5 runs due to runtime variability. Offline includes vector generation, indexing, and training times where applicable; Online is total query search time.}
\label{tab:embedding-time}
\end{table*}
\subsection{Evidence of Benchmark Limitations}

The most compelling evidence for our benchmark limitation hypotheses emerges from the unexpectedly strong performance of simple baselines. On partitioning-based benchmarks (\textsc{Tus}$_{\text{Small}}$, \textsc{Tus}$_{\text{Large}}$, \textsc{Santos}), lexical methods achieve near-perfect precision, matching or exceeding sophisticated models at a fraction of the cost. This directly validates our overlap issue hypothesis: the high schema and value overlap (Figure~\ref{fig:benchmark_overlap}) creates trivial signals that simple lexical matching can exploit. While advanced methods like Starmie or HEARTS also achieve high scores here, the fact that much simpler, non-semantic methods perform nearly identically leads us to conclude that the benchmark itself does not effectively differentiate methods based on deep semantic understanding. This phenomenon, where simpler approaches can achieve comparable or even better results than more complex counterparts, especially when computational costs are considered, has also been observed in related data lake tasks such as table augmentation via join search \cite{cappuzzo_retrieve_2024}.

For \textsc{Pylon},  a different pattern emerges: lexical methods perform considerably worse due to the much lower exact overlap, but general-purpose semantic embeddings excel. SBERT variants, particularly SBERT(V+C) combining column and value information, outperform specialized SOTA models like Starmie. This confirms our general semantics hypothesis that these benchmarks employ vocabulary well-represented in standard pre-trained embeddings, diminishing the advantage of specialized tabular architectures for the TUS task.

\textsc{LB-OpenData} and \textsc{LB-Webtable} exhibit both limitations despite their scale. Simple lexical methods remain surprisingly competitive, while SBERT variants consistently outperform specialized models. The computational demands of sophisticated models create additional practical barriers: Starmie requires substantial offline costs (training and inference) plus over 16 hours to process the queries on the truncated \textsc{LB-OpenData}, and over 21 hours to evaluate the queries of \textsc{LB-Webtable}. HEARTS performs better computationally by leveraging a pre-trained checkpoint without additional training, resulting in a shorter offline processing time, but still under-performs SBERT variants.

\subsection{Ground Truth Reliability Issues} \label{sec:ground_truth_issues}

A notable observation across \textsc{Ugen} and \textsc{LakeBench} derivatives is the significant gap between the $\text{R@k}$ achieved by all methods and the IDEAL recall (Table~\ref{tab:embedding-metrics}). This discrepancy led us to question the reliability of the benchmarks' ground truth labels. We hypothesized that such gaps might indicate not only the limitations of the search methods or the inherent difficulty of the benchmarks but also potential incompleteness or inaccuracies within the ground truth itself. Examining discrepancies at small values of $k$ is particularly revealing, as this scrutinizes the highest-confidence predictions of a system. If a high-performing method frequently disagrees with the ground truth at these top ranks, it may signal issues with the ground truth labels.

To investigate this, we defined two heuristic metrics designed to help identify potential ground truth flaws. Let $\mathcal{Q} = \{Q_1, \ldots, Q_N\}$ be $N$ query tables. For $Q_i \in \mathcal{Q}$, $C_{Q_i, k}$ is the set of top-$k$ candidates retrieved by a search method for $Q_i$, and $G_{Q_i}$ is the set of ground truth candidates labeled unionable with $Q_i$.

\begin{enumerate}
    \item \textbf{GTFP@k (Ground Truth False Positive Rate)}: This measures the fraction of top-$k$ candidates retrieved by a search method that are not labeled as unionable in the original ground truth. A high GTFP@k, especially at small $k$, suggests the method might be identifying valid unionable tables missing from the ground truth, thereby helping us pinpoint its possible \textit{incompleteness}. It is calculated as:
    $$\frac{\sum_{i=1}^{N} |C_{Q_i, k} \setminus G_{Q_i}|}{N \cdot k} $$
    Here, $|C_{Q_i, k} \setminus G_{Q_i}|$ counts retrieved candidates for $Q_i$ that are absent from its ground truth set $G_{Q_i}$. The denominator is the total top-$k$ slots considered across all queries.

    \item \textbf{GTFN@k (Ground Truth False Negative Rate)}: This quantifies the fraction of items labeled as positives in the ground truth that a well-performing search method fails to retrieve within its top-$k$ results (considering a capped expectation up to $k$ items per query). It is calculated as:
    $$\frac{\sum_{i=1}^{N} (\min(k, |G_{Q_i}|) - |G_{Q_i} \cap C_{Q_i, k}|)}{\sum_{i=1}^{N} \min(k, |G_{Q_i}|)} $$
    The term $\min(k, |G_{Q_i}|)$ represents the capped ideal number of ground truth items we would expect to find in the top $k$ for $Q_i$. The numerator sums the "misses" for each query: the difference between this capped ideal and the number of ground truth items actually retrieved. The denominator sums this capped ideal across all queries.
    A high GTFN@k at small $k$ is particularly insightful when investigating ground truth integrity. If we trust the method's ability to discern relevance, a high GTFN@k implies that the method correctly deprioritizes items that, despite being in the ground truth, might be less relevant or even incorrectly labeled as positive. Thus, it can signal potential \textit{incorrectness} within the ground truth. GTFN@k is equivalent to "$1 - \text{CappedRecall@k}$" \cite{thakur2021beir}.
\end{enumerate}

These metrics assume discrepancies between a strong search method and the ground truth may indicate flaws in the latter. While not highly accurate, they helped us identify a smaller, focused subset of query-candidate pairs with disagreements for deeper manual or LLM-based inspection. Results are shown in Table~\ref{tab:gtfp-gtfn-rates-combined}.

Beyond heuristic metrics, we also conduct a more direct--though still imperfect--assessment of \textsc{Ugen}’s ground truth using an LLM-as-a-judge approach. While this method may not capture the same conflicts identified by the cheaper GTFP/GTFN heuristics, it provides a complementary perspective that can offer more precise insights in certain cases. We use \texttt{gemini-2.0-flash-thinking-exp-01-21}\footnote{\href{https://deepmind.google/technologies/gemini/flash-thinking/}{Gemini 2.0 Flash Thinking Model Card}}, chosen for its 1M-token context window, baked-in reasoning abilities, and low hallucination rate\footnote{\href{https://github.com/vectara/hallucination-leaderboard}{Vectara Hallucination Leaderboard}}. This LLM-as-a-judge approach has become increasingly common in recent works~\cite{gu2024survey,wolff2025well}. We gave the LLM both tables in each query-candidate pair, along with a detailed prompt including curated unionable and non-unionable examples from \textsc{Ugen} (see Appendix~\ref{appendix:prompt}) to condition the LLM's understanding of unionability based on the benchmark. Each pair was evaluated in 5 independent runs with \texttt{temperature=0.1}. A sample of 20 LLM outputs was manually validated and showed strong alignment with human judgment. Comparison with original \textsc{Ugen} labels (Table~\ref{tab:llm-gt-comparison}) revealed substantial inconsistencies. Our manual inspection (Appendix~\ref{appendix:ugen_examples}) suggested the LLM often provided more accurate assessments, indicating notable noise in the original ground truth.

Given the scale of \textsc{LB-OpenData} and \textsc{LB-Webtable}, full LLM adjudication was impractical. Instead, we used SBERT(V+C) as our reference search method to compute GTFP@k, focusing on top-ranked pairs not labeled as unionable in the ground truth. As shown in Table~\ref{tab:gtfp-gtfn-rates-combined}, such cases were frequent even at top ranks ($2< k < 5$). To assess ground truth completeness, we manually inspected 20 randomly sampled top-2 and top-3 disagreements. Of these, 19 were genuinely unionable but missing from the ground truth; the remaining pair was correctly non-unionable, with SBERT likely misled by its numeric-only columns. These results suggest non-negligible \textit{incompleteness} in the \textsc{LakeBench} ground truth. Example cases are shown in Appendix~\ref{appendix:lb_examples}.

\begin{table}[htpb]
\centering
\footnotesize
\setlength{\tabcolsep}{2pt}
\begin{tabular}{lccccc}
\hline
\textbf{Benchmark (Metric)} & \textbf{@1} & \textbf{@2} & \textbf{@3} & \textbf{@4} & \textbf{@5} \\
\hline
\textsc{UGEN\(_{\text{V1}}\)} (GTFP) & 0.160 & 0.210 & 0.247 & 0.275 & 0.308 \\
\textsc{UGEN\(_{\text{V1}}\)} (GTFN) & 0.160 & 0.210 & 0.247 & 0.275 & 0.308 \\
\hline
\textsc{UGEN\(_{\text{V2}}\)} (GTFP) & 0.060 & 0.080 & 0.093 & 0.140 & 0.156 \\
\textsc{UGEN\(_{\text{V2}}\)} (GTFN) & 0.060 & 0.080 & 0.093 & 0.140 & 0.156 \\
\hline
\textsc{LB-OpenData} (GTFP) & 0.000 & 0.059 & 0.092 & 0.123 & 0.154 \\
\textsc{LB-OpenData} (GTFN) & 0.000 & 0.054 & 0.080 & 0.105 & 0.132 \\
\hline
\textsc{LB-Webtable} (GTFP) & 0.000 & 0.110 & 0.198 & 0.296 & 0.377 \\
\textsc{LB-Webtable} (GTFN) & 0.000 & 0.110 & 0.197 & 0.295 & 0.376 \\
\hline
\end{tabular}
\caption{Disagreement rates of top-$k$ retrieved results between SBERT and the ground truth across different benchmarks. For \textsc{Ugen},  the query table is not considered a candidate to itself, so values at @1 reflect actual disagreement. For \textsc{LakeBench} variants, the ground truth is normalized to include the query table as a valid candidate for itself. Therefore, the top-1 match is always correct by construction, yielding no disagreement @1.}
\label{tab:gtfp-gtfn-rates-combined}
\end{table}

\begin{table}[htpb]
\centering
\footnotesize
\setlength{\tabcolsep}{4pt}
\begin{tabular}{llcc}
\hline
\textbf{GT Label} & \textbf{LLM Judge} & \textbf{UGEN V1} & \textbf{UGEN V2} \\
\hline
Unionable & Non-unionable & 24.8\% & 0.0\% \\
Non-unionable & Unionable & 33.8\% & 23.6\% \\
Non-unionable & Non-unionable & 16.2\% & 76.4\% \\
Unionable & Non-unionable & 25.2\% & 0.0\% \\
\hline
\end{tabular}
\caption{Breakdown of agreement and disagreement between ground truth labels and LLM-based judgments.}
\label{tab:llm-gt-comparison}
\end{table}

In summary, our investigations, combining heuristic metrics, LLM-based adjudication, and manual inspection, reveal the presence of non-negligible noise and incompleteness within the original benchmark labels for both \textsc{Ugen} and \textsc{LakeBench}. Consequently, performance metrics reported on these benchmarks may be influenced by these underlying ground truth issues, potentially misrepresenting true task difficulty or method capabilities.

\subsection{Implications for Measuring Progress}

Our experiments reveal several critical issues. Benchmark scores often fail to measure true semantic capabilities, as simple lexical or general embedding methods can match or outperform specialized models by exploiting excessive domain overlap, semantic simplicity, or ground truth inconsistency. This suggests that current benchmarks may inadvertently reward adaptation to these characteristics, making it difficult to quantify the practical benefits of progress on sophisticated TUS methods capabilities within these settings. These persistent issues also point to a fundamental challenge, the lack of a precise, operational definition for unionability, mirroring broader difficulties in dataset search~\citep{hulsebos_it_2024} and highlighting the need to address the subjective, context-dependent nature of table compatibility in practice.

\section{Towards Better TUS Benchmarks} \label{sec:recommendations}

In industry practice, unionability judgments are inherently subjective, depending on analytical goals, domain contexts, data accessibility constraints \cite{martorana2025metadatadriventableunionsearch}, and user preferences \cite{DBLP:conf/vldb/MirzaeiR23}. Yet current benchmarks impose fixed definitions, creating a disconnect with practical utility: methods excelling on benchmarks often falter in real-world scenarios demanding different compatibility thresholds. Addressing this requires benchmark designs that embrace contextual variability and provide a stable foundation for evaluation, lest even advanced methods fall short in practice.

\paragraph{Rethinking Benchmark Design Principles:}
Overcoming current benchmark limitations requires a shift in design focusing on three key principles: (1) actively reducing artifactual overlap while introducing controlled semantic heterogeneity to better reflect real-world schema and value divergence; (2) incorporating realistic domain complexity beyond general vocabularies, addressing challenges like non-descriptive schemas and proprietary terms where LLMs struggle~\citep{bodensohn2025unveilingchallengesllmsenterprise}, thus emphasizing domain-specific training that may require industry collaboration; and (3) rethinking ground truth representation by replacing brittle binary labels with richer, nuanced formats validated through multi-stage adjudication to improve completeness and consistency.

\paragraph{Exploring Implementation Pathways:}

Translating these principles into practice requires concrete strategies for benchmark design and evaluation. One approach is to develop (1) scenario-driven micro-benchmarks targeting specific challenges such as schema drift simulation or value representation noise, enabling more granular analysis than coarse end-to-end metrics. Another is (2) advancing controllable synthetic data generation, following LLM-based methods like \textsc{Ugen} \citep{pal2024alt}, to verifiably embed semantic constraints or domain knowledge, supporting diverse testbeds when real data is unavailable or sensitive. Equally important is (3) exploring adaptive, interactive evaluation frameworks such as human-in-the-loop systems, which would dynamically adjust relevance criteria based on user feedback to better capture the subjective nature of unionability. Tools like LakeVisage \citep{hu2025lakevisage} further enhance usability and trust by recommending visualizations that help users interpret relationships among returned tables, improving transparency and interpretability in union search systems.Incorporating natural language preferences is also key. The recent \textsc{NLCTables} benchmark \citep{cui2025nlctablesdatasetmarryingnatural} advances this by introducing NL conditions for union and join searches on column values and table size constraints. However, its predicate-style conditions may be better addressed via post-retrieval filtering (e.g., translating NL to SQL predicates with an LLM), avoiding early discard of unionable candidates and unnecessary retrieval model complexity. To drive further advancement, benchmarks should incorporate (4) natural language conditions that capture key aspects of unionability and joinability, including specifications about the characteristics of the final integrated table or conditional integration logic. For example, a challenging predicate might require identifying tables that can be "joined with a query table on column A, unioned on columns B and C, and also contain an additional column D providing specific contextual information about a particular attribute." Such conditions would demand deeper reasoning capabilities from data integration systems and encourage the development of more sophisticated methods for Table Union and Join Search. Finally, moving beyond binary success metrics, future benchmarks could adopt (5) multi-faceted evaluation frameworks using richer ground truth representations to assess unionability across dimensions like schema compatibility, semantic type alignment, value distribution similarity, and task-specific relevance, offering a more holistic evaluation than current standards.

\section{Conclusion} \label{sec:conclusion}

Our analysis of TUS benchmarks highlights three major limitations: excessive overlap in partitioning-based datasets, semantics easily captured by pre-trained embeddings, and non-negligible ground-truth inconsistencies. The first two allow simple baselines to rival sophisticated models with far lower computational cost, showing that high performance isn’t necessarily tied to advanced semantic reasoning. The third undermines evaluation validity, as scores may reflect misalignment with flawed ground truth rather than actual benchmark difficulty. This gap between benchmark performance and true semantic capability suggests current evaluations often reward adaptation to benchmark-specific artifacts. To address this, we propose design principles that better reflect the complex, subjective nature of real-world table union search.

\paragraph{Limitations:} Our study examined selected benchmarks and methods, with broader evaluation potentially revealing more insight. Our investigation of ground truth issues in \textsc{Ugen} and \textsc{LakeBench}, while systematic, identifies certain patterns without exhaustive quantification.

\paragraph{Future Work:} Developing benchmarks aligned with our proposed criteria represents the next step towards ensuring that measured progress translates to meaningful real-world utility.

\bibliography{anthology}

\begin{thebibliography}{33}
\providecommand{\natexlab}[1]{#1}

\bibitem[{Bodensohn et~al.(2025)Bodensohn, Brackmann, Vogel, Sanghi, and Binnig}]{bodensohn2025unveilingchallengesllmsenterprise}
Jan-Micha Bodensohn, Ulf Brackmann, Liane Vogel, Anupam Sanghi, and Carsten Binnig. 2025.
\newblock \href {https://arxiv.org/abs/2504.10950} {Unveiling challenges for llms in enterprise data engineering}.
\newblock \emph{Preprint}, arXiv:2504.10950.

\bibitem[{Bogatu et~al.(2020)Bogatu, Fernandes, Paton, and Konstantinou}]{bogatu_d3l_2020}
Alex Bogatu, Alvaro A.~A. Fernandes, Norman~W. Paton, and Nikolaos Konstantinou. 2020.
\newblock \href {https://doi.org/10.1109/ICDE48307.2020.00067} {{D3L}: {Dataset} {Discovery} in {Data} {Lakes}}.
\newblock In \emph{2020 {IEEE} 36th {International} {Conference} on {Data} {Engineering} ({ICDE})}, pages 709--720.
\newblock ArXiv:2011.10427 [cs].

\bibitem[{Boutaleb et~al.(2025)Boutaleb, Almutawa, Amann, Angarita, and Naacke}]{boutaleb2025hearts}
Allaa Boutaleb, Alaa Almutawa, Bernd Amann, Rafael Angarita, and Hubert Naacke. 2025.
\newblock \href {https://openreview.net/forum?id=XgRbxO9pLJ} {{HEARTS}: Hypergraph-based related table search}.
\newblock In \emph{ELLIS workshop on Representation Learning and Generative Models for Structured Data}.

\bibitem[{Cappuzzo et~al.(2024)Cappuzzo, Varoquaux, Coelho, and Papotti}]{cappuzzo_retrieve_2024}
Riccardo Cappuzzo, Gaël Varoquaux, Aimee Coelho, and Paolo Papotti. 2024.
\newblock \href {https://doi.org/10.48550/arXiv.2402.06282} {Retrieve, merge, predict: Augmenting tables with data lakes}.
\newblock \emph{CoRR}, abs/2402.06282.

\bibitem[{Castelo et~al.(2021)Castelo, Rampin, Santos, Bessa, Chirigati, and Freire}]{castelo_auctus_2021-1}
Sonia Castelo, Rémi Rampin, Aécio Santos, Aline Bessa, Fernando Chirigati, and Juliana Freire. 2021.
\newblock \href {https://doi.org/10.14778/3476311.3476346} {Auctus: a dataset search engine for data discovery and augmentation}.
\newblock \emph{Proceedings of the VLDB Endowment}, 14(12):2791--2794.

\bibitem[{Chen et~al.(2023)Chen, Sarkar, Lausen, Srinivasan, Zha, Huang, and Karypis}]{chen_hytrel_2023}
Pei Chen, Soumajyoti Sarkar, Leonard Lausen, Balasubramaniam Srinivasan, Sheng Zha, Ruihong Huang, and George Karypis. 2023.
\newblock Hytrel: Hypergraph-enhanced tabular data representation learning.
\newblock \emph{Advances in Neural Information Processing Systems}, 36:32173--32193.

\bibitem[{Cong et~al.(2023)Cong, Nargesian, and Jagadish}]{cong_pylon_2023}
Tianji Cong, Fatemeh Nargesian, and H.~V. Jagadish. 2023.
\newblock Pylon: Semantic table union search in data lakes.
\newblock \emph{CoRR}, abs/2301.04901.

\bibitem[{Cui et~al.(2025)Cui, Li, Chen, Shou, and Chen}]{cui2025nlctablesdatasetmarryingnatural}
Lingxi Cui, Huan Li, Ke~Chen, Lidan Shou, and Gang Chen. 2025.
\newblock \href {https://doi.org/10.48550/ARXIV.2504.15849} {Nlctables: {A} dataset for marrying natural language conditions with table discovery}.
\newblock \emph{CoRR}, abs/2504.15849.

\bibitem[{Deng et~al.(2024)Deng, Chai, Cao, Yuan, Chen, Yu, Sun, Wang, Li, Cao, Jin, Zhang, Jiang, Zhang, Wang, Yuan, Wang, and Tang}]{deng_lakebench_2024}
Yuhao Deng, Chengliang Chai, Lei Cao, Qin Yuan, Siyuan Chen, Yanrui Yu, Zhaoze Sun, Junyi Wang, Jiajun Li, Ziqi Cao, Kaisen Jin, Chi Zhang, Yuqing Jiang, Yuanfang Zhang, Yuping Wang, Ye~Yuan, Guoren Wang, and Nan Tang. 2024.
\newblock \href {https://doi.org/10.14778/3659437.3659448} {{LakeBench}: {A} {Benchmark} for {Discovering} {Joinable} and {Unionable} {Tables} in {Data} {Lakes}}.
\newblock \emph{Proceedings of the VLDB Endowment}, 17(8):1925--1938.

\bibitem[{Devlin et~al.(2019)Devlin, Chang, Lee, and Toutanova}]{devlin-etal-2019-bert}
Jacob Devlin, Ming-Wei Chang, Kenton Lee, and Kristina Toutanova. 2019.
\newblock \href {https://doi.org/10.18653/v1/N19-1423} {{BERT}: Pre-training of deep bidirectional transformers for language understanding}.
\newblock In \emph{Proceedings of the 2019 Conference of the North {A}merican Chapter of the Association for Computational Linguistics: Human Language Technologies, Volume 1 (Long and Short Papers)}, pages 4171--4186, Minneapolis, Minnesota. Association for Computational Linguistics.

\bibitem[{Douze et~al.(2024)Douze, Guzhva, Deng, Johnson, Szilvasy, Mazaré, Lomeli, Hosseini, and Jégou}]{douze2024faiss}
Matthijs Douze, Alexandr Guzhva, Chengqi Deng, Jeff Johnson, Gergely Szilvasy, Pierre-Emmanuel Mazaré, Maria Lomeli, Lucas Hosseini, and Hervé Jégou. 2024.
\newblock \href {https://arxiv.org/abs/2401.08281} {The faiss library}.

\bibitem[{Fan et~al.(2023{\natexlab{a}})Fan, Wang, Li, and Miller}]{fan_table_2023}
Grace Fan, Jin Wang, Yuliang Li, and Renée~J. Miller. 2023{\natexlab{a}}.
\newblock \href {https://doi.org/10.1145/3555041.3589409} {Table {Discovery} in {Data} {Lakes}: {State}-of-the-art and {Future} {Directions}}.
\newblock In \emph{Companion of the 2023 {International} {Conference} on {Management} of {Data}}, pages 69--75, Seattle WA USA. ACM.

\bibitem[{Fan et~al.(2023{\natexlab{b}})Fan, Wang, Li, Zhang, and Miller}]{fan_starmie_2023}
Grace Fan, Jin Wang, Yuliang Li, Dan Zhang, and Ren{\'{e}}e~J. Miller. 2023{\natexlab{b}}.
\newblock \href {https://doi.org/10.14778/3587136.3587146} {Semantics-aware dataset discovery from data lakes with contextualized column-based representation learning}.
\newblock \emph{Proc. {VLDB} Endow.}, 16(7):1726--1739.

\bibitem[{Gomm and Hulsebos(2025)}]{gomm2025metadata}
Daniel Gomm and Madelon Hulsebos. 2025.
\newblock \href {https://openreview.net/forum?id=rELWIvq2Qy} {Metadata matters in dense table retrieval}.
\newblock In \emph{ELLIS workshop on Representation Learning and Generative Models for Structured Data}.

\bibitem[{Gu et~al.(2024)Gu, Jiang, Shi, Tan, Zhai, Xu, Li, Shen, Ma, Liu, Wang, and Guo}]{gu2024survey}
Jiawei Gu, Xuhui Jiang, Zhichao Shi, Hexiang Tan, Xuehao Zhai, Chengjin Xu, Wei Li, Yinghan Shen, Shengjie Ma, Honghao Liu, Yuanzhuo Wang, and Jian Guo. 2024.
\newblock \href {https://doi.org/10.48550/ARXIV.2411.15594} {A survey on llm-as-a-judge}.
\newblock \emph{CoRR}, abs/2411.15594.

\bibitem[{Hu et~al.(2023)Hu, Wang, Qin, Lei, Shen, Faloutsos, Katsifodimos, Karypis, Wen, and Yu}]{hu_autotus_2023}
Xuming Hu, Shen Wang, Xiao Qin, Chuan Lei, Zhengyuan Shen, Christos Faloutsos, Asterios Katsifodimos, George Karypis, Lijie Wen, and Philip~S. Yu. 2023.
\newblock \href {https://doi.org/10.18653/v1/2023.findings-acl.233} {{AUTOTUS}: {Automatic} {Table} {Union} {Search} with {Tabular} {Representation} {Learning}}.
\newblock In \emph{Findings of the {Association} for {Computational} {Linguistics}: {ACL} 2023}, pages 3786--3800, Toronto, Canada. Association for Computational Linguistics.

\bibitem[{Hu et~al.(2025)Hu, Wang, and Rahman}]{hu2025lakevisage}
Yihao Hu, Jin Wang, and Sajjadur Rahman. 2025.
\newblock \href {https://doi.org/10.48550/ARXIV.2504.02150} {Lakevisage: Towards scalable, flexible and interactive visualization recommendation for data discovery over data lakes}.
\newblock \emph{CoRR}, abs/2504.02150.

\bibitem[{Hulsebos et~al.(2023)Hulsebos, Demiralp, and Groth}]{hulsebos2023gittables}
Madelon Hulsebos, {\c{C}}agatay Demiralp, and Paul Groth. 2023.
\newblock \href {https://doi.org/10.1145/3588710} {Gittables: {A} large-scale corpus of relational tables}.
\newblock \emph{Proc. {ACM} Manag. Data}, 1(1):30:1--30:17.

\bibitem[{Hulsebos et~al.(2024)Hulsebos, Lin, Shankar, and Parameswaran}]{hulsebos_it_2024}
Madelon Hulsebos, Wenjing Lin, Shreya Shankar, and Aditya Parameswaran. 2024.
\newblock \href {https://doi.org/10.1145/3665939.3665959} {It {Took} {Longer} than {I} was {Expecting}: {Why} is {Dataset} {Search} {Still} so {Hard}?}
\newblock In \emph{Proceedings of the 2024 {Workshop} on {Human}-{In}-the-{Loop} {Data} {Analytics}}, pages 1--4, Santiago AA Chile. ACM.

\bibitem[{Khatiwada et~al.(2023)Khatiwada, Fan, Shraga, Chen, Gatterbauer, Miller, and Riedewald}]{khatiwada_santos_2023}
Aamod Khatiwada, Grace Fan, Roee Shraga, Zixuan Chen, Wolfgang Gatterbauer, Renée~J. Miller, and Mirek Riedewald. 2023.
\newblock \href {https://doi.org/10.1145/3588689} {{SANTOS}: {Relationship}-based {Semantic} {Table} {Union} {Search}}.
\newblock \emph{Proceedings of the ACM on Management of Data}, 1(1):1--25.

\bibitem[{Khatiwada et~al.(2025)Khatiwada, Kokel, Abdelaziz, Chaudhury, Dolby, Hassanzadeh, Huang, Pedapati, Samulowitz, and Srinivas}]{khatiwada_tabsketchfm_2024}
Aamod Khatiwada, Harsha Kokel, Ibrahim Abdelaziz, Subhajit Chaudhury, Julian Dolby, Oktie Hassanzadeh, Zhenhan Huang, Tejaswini Pedapati, Horst Samulowitz, and Kavitha Srinivas. 2025.
\newblock Tabsketchfm: Sketch-based tabular representation learning for data discovery over data lakes.
\newblock \emph{IEEE ICDE}.

\bibitem[{Martorana et~al.(2025)Martorana, Kuhn, and van Ossenbruggen}]{martorana2025metadatadriventableunionsearch}
Margherita Martorana, Tobias Kuhn, and Jacco van Ossenbruggen. 2025.
\newblock \href {https://doi.org/10.48550/ARXIV.2502.20945} {Metadata-driven table union search: Leveraging semantics for restricted access data integration}.
\newblock \emph{CoRR}, abs/2502.20945.

\bibitem[{McInnes et~al.(2017)McInnes, Healy, Astels et~al.}]{mcinnes2017hdbscan}
Leland McInnes, John Healy, Steve Astels, and 1 others. 2017.
\newblock hdbscan: Hierarchical density based clustering.
\newblock \emph{J. Open Source Softw.}, 2(11):205.

\bibitem[{McInnes et~al.(2018)McInnes, Healy, Saul, and Gro{\ss}berger}]{mcinnes2018umap}
Leland McInnes, John Healy, Nathaniel Saul, and Lukas Gro{\ss}berger. 2018.
\newblock \href {https://doi.org/10.21105/JOSS.00861} {{UMAP:} uniform manifold approximation and projection}.
\newblock \emph{J. Open Source Softw.}, 3(29):861.

\bibitem[{Mirzaei and Rafiei(2023)}]{DBLP:conf/vldb/MirzaeiR23}
Hamed Mirzaei and Davood Rafiei. 2023.
\newblock \href {https://ceur-ws.org/Vol-3462/TADA2.pdf} {Table union search with preferences}.
\newblock In \emph{Joint Proceedings of Workshops at the 49th International Conference on Very Large Data Bases {(VLDB} 2023), Vancouver, Canada, August 28 - September 1, 2023}, volume 3462 of \emph{{CEUR} Workshop Proceedings}. CEUR-WS.org.

\bibitem[{Nargesian et~al.(2018)Nargesian, Zhu, Pu, and Miller}]{nargesian_tus_2018}
Fatemeh Nargesian, Erkang Zhu, Ken~Q. Pu, and Renée~J. Miller. 2018.
\newblock \href {https://doi.org/10.14778/3192965.3192973} {{TUS}: {Table} union search on open data}.
\newblock \emph{Proceedings of the VLDB Endowment}, 11(7):813--825.

\bibitem[{Pal et~al.(2024)Pal, Khatiwada, Shraga, and Miller}]{pal2024alt}
Koyena Pal, Aamod Khatiwada, Roee Shraga, and Ren{\'e}e~J Miller. 2024.
\newblock Alt-gen: Benchmarking table union search using large language models.
\newblock \emph{Proceedings of the VLDB Endowment. ISSN}, 2150:8097.

\bibitem[{Pellissier~Tanon et~al.(2020)Pellissier~Tanon, Weikum, and Suchanek}]{yago4}
Thomas Pellissier~Tanon, Gerhard Weikum, and Fabian Suchanek. 2020.
\newblock Yago 4: A reason-able knowledge base.
\newblock In \emph{The Semantic Web}, pages 583--596, Cham. Springer International Publishing.

\bibitem[{Reimers and Gurevych(2019)}]{reimers-2019-sentence-bert}
Nils Reimers and Iryna Gurevych. 2019.
\newblock \href {https://doi.org/10.18653/v1/D19-1410} {Sentence-{BERT}: Sentence embeddings using {S}iamese {BERT}-networks}.
\newblock In \emph{Proceedings of the 2019 Conference on Empirical Methods in Natural Language Processing and the 9th International Joint Conference on Natural Language Processing (EMNLP-IJCNLP)}, pages 3982--3992, Hong Kong, China. Association for Computational Linguistics.

\bibitem[{Sarma et~al.(2012)Sarma, Fang, Gupta, Halevy, Lee, Wu, Xin, and Yu}]{sarma2012finding}
Anish~Das Sarma, Lujun Fang, Nitin Gupta, Alon~Y Halevy, Hongrae Lee, Fei Wu, Reynold Xin, and Cong Yu. 2012.
\newblock Finding related tables.
\newblock In \emph{SIGMOD Conference}, volume~10, pages 2213836--2213962.

\bibitem[{Thakur et~al.(2021)Thakur, Reimers, R{\"u}ckl{\'e}, Srivastava, and Gurevych}]{thakur2021beir}
Nandan Thakur, Nils Reimers, Andreas R{\"u}ckl{\'e}, Abhishek Srivastava, and Iryna Gurevych. 2021.
\newblock \href {https://openreview.net/forum?id=wCu6T5xFjeJ} {{BEIR}: A heterogeneous benchmark for zero-shot evaluation of information retrieval models}.
\newblock In \emph{Thirty-fifth Conference on Neural Information Processing Systems Datasets and Benchmarks Track (Round 2)}.

\bibitem[{Wolff and Hulsebos(2025)}]{wolff2025well}
Cornelius Wolff and Madelon Hulsebos. 2025.
\newblock How well do llms reason over tabular data, really?
\newblock \emph{arXiv preprint arXiv:2505.07453}.

\bibitem[{Zhu et~al.(2016)Zhu, Nargesian, Pu, and Miller}]{zhu_lsh_2016}
Erkang Zhu, Fatemeh Nargesian, Ken~Q. Pu, and Ren{\'{e}}e~J. Miller. 2016.
\newblock \href {https://doi.org/10.14778/2994509.2994534} {{LSH} ensemble: Internet-scale domain search}.
\newblock \emph{Proc. {VLDB} Endow.}, 9(12):1185--1196.

\end{thebibliography}

\newpage 
\appendix

\section{Benchmark Overlap}\label{appendix:overlap}

\begin{figure*}[htbp] 
    \centering
    \includegraphics[width=1.0\linewidth]{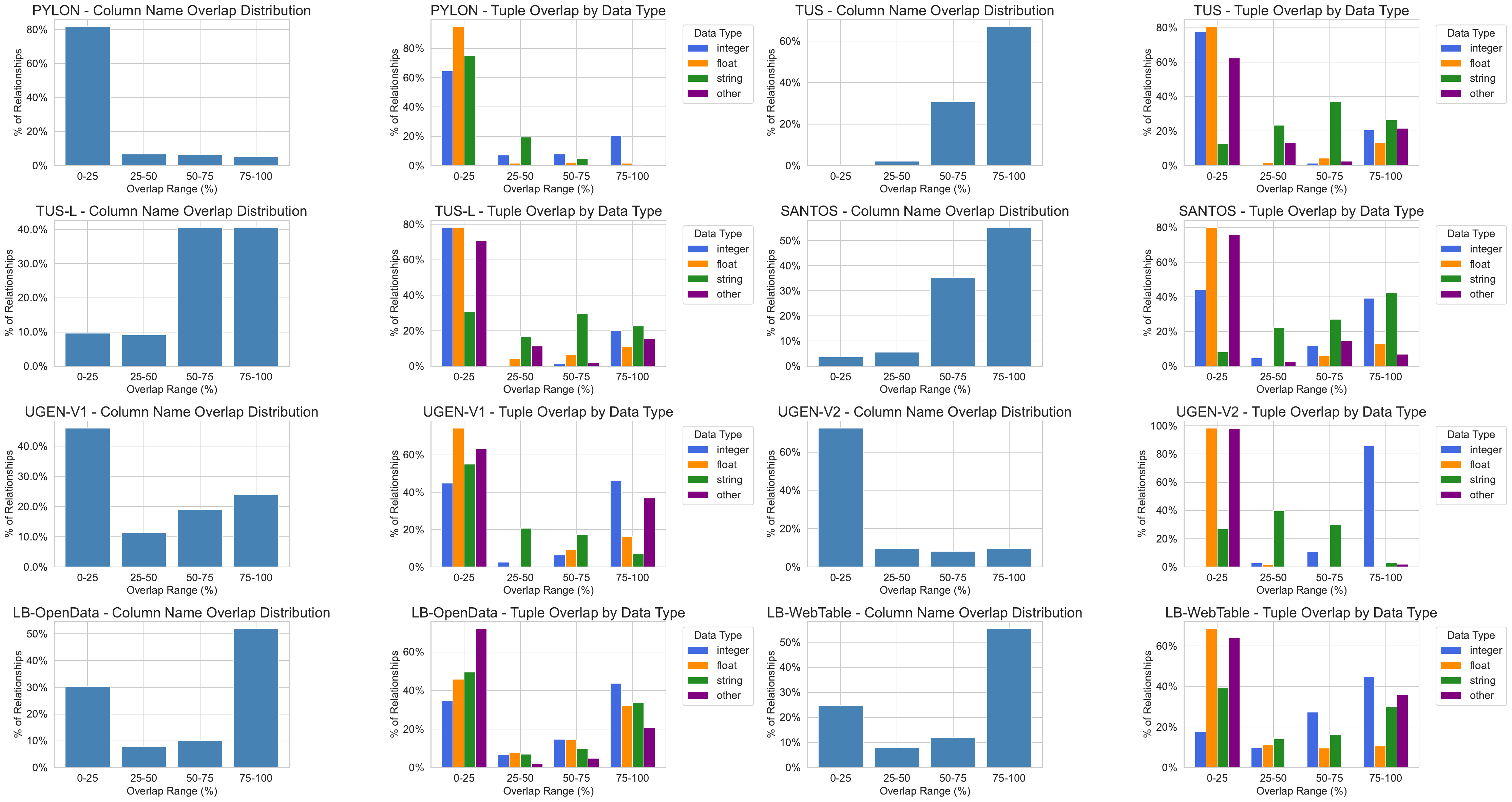} 
    \caption{Distribution of exact column name and tuple overlap across different benchmarks, broken down by data type (String, Numeric, Datetime, Other). Each subplot represents a benchmark, showing the percentage of ground truth pairs falling into different overlap ranges.} 
    \label{fig:overlap_by_datatype} 
\end{figure*}

As discussed in section~\ref{sec:overlap_artifact}, the degree of lexical overlap (both in column names and values) between query and candidate tables in benchmark ground truths can significantly influence model performance. Methods sensitive to surface-level similarity might perform well on benchmarks with high overlap without necessarily capturing deeper semantic relationships. This section provides a more detailed breakdown of overlap coefficients by data type across the different benchmarks evaluated. Figure~\ref{fig:overlap_by_datatype} presents these distributions.

\section{Implementation and Evaluation Details} \label{appendix:exp_setup}

This appendix provides supplementary details regarding the implementation of baseline methods, SOTA models, and the evaluation procedure used in our experiments, complementing the core methodology described in Sections \ref{sec:baselines_strategy} and \ref{sec:eval_procedure}.

\subsection{Lexical Baselines (Hashing, TF-IDF, Count) Implementation Details} \label{appendix:baseline_impl}

\paragraph{Vectorizers:} We used implementations from scikit-learn\footnote{\url{https://scikit-learn.org/stable/api/sklearn.feature_extraction.html}}. All vectorizers were configured with \texttt{lowercase=True}.
        \begin{itemize}
            \item \texttt{TfidfVectorizer} and \texttt{CountVectorizer}: Used an \texttt{ngram\_range=(1, 2)}. Their vocabulary was constructed by first collecting unique tokens from all columns across the entire corpus (query tables included), ensuring a consistent feature space.
            \item \texttt{HashingVectorizer}: Used an \texttt{ngram\_range=(1, 1)} and \texttt{alternate\_sign=False}.
        \end{itemize}
\paragraph{Input Data:} For each table, we randomly sampled up to 1000 unique non-null cell values per column.

\paragraph{Vectorization:} Each column's sampled values were treated as a document and vectorized into a 4096-dimensional vector using the appropriately fitted or configured vectorizer.

\subsection{SOTA Method Implementation Details} \label{appendix:sota_impl}

\subsubsection{Starmie:} \footnote{\href{https://github.com/megagonlabs/starmie}{Starmie GitHub Repository}}
We utilized the implementation and recommendations from the original Starmie paper \citep{fan_starmie_2023}.
\paragraph{Training Setup:} The provided RoBERTa-based model was retrained for 10 epochs on each benchmark. Key hyperparameters included: batch size 32, projection dimension 768, learning rate 5e-5, max sequence length 256, and fp16 precision.
\paragraph{Sampling and Augmentation Strategies:} Starmie employs specific strategies during contrastive pre-training to generate positive pairs (views of the same column). The strategies, based on the definitions in the original paper, are:
        \begin{itemize}[noitemsep, topsep=0pt, partopsep=0pt, itemsep=1pt]
            \item \textit{TF-IDF Entity Sampling (`tfidf\_entity`):} Samples cells in columns that have the highest average TF-IDF scores calculated over their tokens.
            \item \textit{Alpha Head Sampling (`alphaHead`):} Samples the first N tokens sorted alphabetically.
            \item \textit{Column Dropping Augmentation (`drop\_col`):} Creates augmented views by dropping a random subset of columns from the table.
            \item \textit{Drop Cell Augmentation (`drop\_cell`):} Creates augmented views by dropping random cells within the table.
        \end{itemize}
        We followed the paper's recommendations for each benchmark, detailed in Table~\ref{tab:starmie_strategies}. For benchmarks not explicitly mentioned in the original paper (\textsc{Pylon},  \textsc{Ugen},  \textsc{LakeBench} derivatives), we applied the same strategies recommended for the \textsc{Santos} benchmark.

\paragraph{Evaluation:} We used the "Pruning" search strategy described in the Starmie paper, also referred to as "bounds" in the original implementation. This involves a maximum bipartite matching approach on a pruned set of candidate column pairs to calculate table similarity, offering higher efficiency compared to naive matching, while remaining more precise than approximate search approaches.

\begin{table}[h]
\centering
\footnotesize
\renewcommand{\arraystretch}{1.1}
\setlength{\tabcolsep}{4pt}
\begin{tabular}{@{}l|l|l@{}}
\toprule
\textbf{Benchmark} & \textbf{Sampling} & \textbf{Augmentation} \\
\midrule
SANTOS & tfidf\_entity & drop\_col \\
TUS (Small) & alphaHead & drop\_cell \\
TUS$_{\text{Large}}$ & tfidf\_entity & drop\_cell \\ 
PYLON & tfidf\_entity & drop\_col \\ 
UGEN$_{\text{V1}}$ & tfidf\_entity & drop\_col \\ 
UGEN$_{\text{V2}}$ & tfidf\_entity & drop\_col \\ 
\textsc{LB-OpenData} & tfidf\_entity & drop\_col \\ 
\textsc{LB-Webtable} & tfidf\_entity & drop\_col \\ 
\bottomrule
\end{tabular}
\caption{Starmie sampling and augmentation strategies applied per benchmark.}
\label{tab:starmie_strategies}
\end{table}

\subsubsection{HEARTS:}\footnote{\href{https://github.com/Allaa-boutaleb/HEARTS}{HEARTS GitHub Repository}}

\paragraph{Model:} Employs pre-trained HyTrel embeddings \citep{chen_hytrel_2023}, utilizing a publicly available checkpoint trained with a contrastive learning objective\footnote{\url{https://github.com/awslabs/hypergraph-tabular-lm/tree/main/checkpoints}}. No further finetuning was performed.
\paragraph{Evaluation Strategy:} We adopted the best-performing search strategy reported in the HEARTS repository for each benchmark:
        \begin{itemize}[noitemsep, topsep=0pt, partopsep=0pt, itemsep=1pt]
            \item \textbf{Cluster Search (for \textsc{Santos},  \textsc{Pylon},  UGEN$_{\text{V1}}$, UGEN$_{\text{V2}}$):} This strategy first reduces the dimensionality of the pre-trained HyTrel column embeddings using UMAP \citep{mcinnes2018umap} and then performs clustering using HDBSCAN \citep{mcinnes2017hdbscan}. Default parameters provided in the HEARTS repository were used for both UMAP and HDBSCAN within this search method. Table similarity is derived based on cluster assignments.
            \item \textbf{FAISS + Max Pooling (for TUS$_{\text{Small}}$, TUS$_{\text{Large}}$, \textsc{LB-OpenData}, \textsc{LB-Webtable}):} This strategy uses FAISS \citep{douze2024faiss} for efficient similarity search. Table vectors are computed by max-pooling the embeddings of their constituent columns before indexing and searching.
        \end{itemize}

\subsection{Hardware}\label{appendix:hardware}
Our experiments were conducted using the following setup:
\begin{itemize}[noitemsep, topsep=0pt, partopsep=0pt, itemsep=2pt]
    \item CPU: Intel Xeon Gold 6330: 4 cores / 8 threads @ 2.00 GHz.
    \item GPU: 40GB MIG partition of NVIDIA A100 (used for SBERT embedding generation and SOTA models training/inference).
    \item 64 Go DDR4 RAM.
\end{itemize}

\section{Inconsistent Ground Truth Examples} \label{appendix:ground_truth_examples}

This section provides illustrative examples of the ground truth inconsistencies identified in the \textsc{Ugen} and \textsc{LakeBench} benchmarks during our analysis (Section \ref{sec:ground_truth_issues}). We categorize these into False Positives (pairs incorrectly labeled as unionable) and False Negatives (pairs incorrectly labeled as non-unionable or missed).

\subsection{UGEN Benchmark Inconsistencies}\label{appendix:ugen_examples}

Figures \ref{fig:ugen_incorrect_positives_app} and \ref{fig:ugen_incorrect_negatives_app} showcase examples from \textsc{Ugen} variants.

\begin{figure}[h] 
    \centering
    \begin{subfigure}{1.0\columnwidth}
        \centering
        \begin{adjustbox}{width=\columnwidth,center}
            \small
            \renewcommand{\arraystretch}{1.1} \setlength{\tabcolsep}{4pt}
            \begin{tabular}{|p{\linewidth}|} \hline
            \makecell[l]{\textbf{Query:} Anthropology\_FGTNBDWF.csv \\ \textbf{Candidate:} Anthropology\_N30U114M.csv} \\ \hline
            \begin{adjustbox}{width=\linewidth}
            \begin{tabular}{@{}ccccccc@{}} \toprule Age & Culture & Arena & Domain & Meaning & Origin & Activity \\ \midrule 1 & Neo. & Arch. & Past & Prim. & Africa & Hunt. \\ 2 & Islam. & Artif. & Hist. & Cplx. & Asia & Farm. \\ \bottomrule \end{tabular}
            \end{adjustbox} \\ \hline
            \begin{adjustbox}{width=\linewidth}
            \begin{tabular}{@{}ccccc@{}} \toprule Artifact & Language & Technology & Education & Society \\ \midrule 1 & English & GPS & Political & Communal \\ 2 & Latin & Smartphone & Scientific & Global \\ \bottomrule \end{tabular}
            \end{adjustbox} \\ \hline
            \end{tabular}
        \end{adjustbox}
        \caption{UGEN$_{\text{V1}}$ Example: Tables discussing structurally and semantically distinct aspects of Anthropology (historical cultures vs. social technology), originally labeled unionable despite conceptual incompatibility.}
        \label{tab:ugen_incorrect_v1_app}
    \end{subfigure}
    \begin{subfigure}{1.0\columnwidth}
        \centering
         \begin{adjustbox}{width=\columnwidth,center}
            \small
            \renewcommand{\arraystretch}{1.1} \setlength{\tabcolsep}{3pt}
            \begin{tabular}{|p{\linewidth}|} \hline
            \makecell[l]{\textbf{Query:} Anthropology\_N7BS08I4.csv \\ \textbf{Candidate:} Anthropology\_VS4SJ2VH.csv} \\ \hline
            \begin{adjustbox}{width=\linewidth}
            \begin{tabular}{@{}llll@{}} \toprule Site Name & Location & Period & Culture \\ \midrule Olduvai Gorge & Tanzania, Africa & Pliocene & Hominin \\ Teotihuacan & Central Mexico & Early Classic & Teotihuacanos \\ \bottomrule \end{tabular}
            \end{adjustbox} \\ \hline
            \begin{adjustbox}{width=\linewidth}
            \begin{tabular}{@{}llll@{}} \toprule Age Group & Clothing & Food & Housing \\ \midrule Children (0-12) & Tunics, hides & Porridge, roots & Huts (branch) \\ Teenagers (13-19) & Garments, beads & Grains, stews & Huts (woven) \\ \bottomrule \end{tabular}
            \end{adjustbox} \\ \hline
            \end{tabular}
         \end{adjustbox}
        \caption{UGEN$_{\text{V2}}$ Example: Tables about archaeological sites versus demographic lifestyles, representing fundamentally different entity types despite the shared Anthropology topic.}
        \label{tab:ugen_incorrect_v2_app}
    \end{subfigure}
    \caption{Examples of \textsc{Ugen} where pairs labeled unionable in the original ground truth exhibit significant semantic/structural divergence suggesting non-unionability.}
    \label{fig:ugen_incorrect_positives_app}
\end{figure}

\begin{figure}[htbp] 
    \centering
    \begin{subfigure}{1.0\columnwidth}
        \centering
        \begin{adjustbox}{width=\columnwidth,center}
            \small
            \renewcommand{\arraystretch}{1.1} \setlength{\tabcolsep}{3pt}
            \begin{tabular}{|p{\linewidth}|} \hline
            \makecell[l]{\textbf{Query:} Archeology\_2LWSQ5A2.csv \\ \textbf{Candidate:} Archeology\_3ML53C0M.csv} \\ \hline
            \begin{adjustbox}{width=\linewidth}
            \begin{tabular}{@{}cccccc@{}} \toprule Discovery & Item & Artifact & Date & Culture & Region \\ \midrule Giza Pyramid & Scroll & Diamond & \textasciitilde2500 BC & Anc. Egypt & N. Africa \\ Tut. Tomb & Knife & Stone Tab. & 1323 BC & Anc. Egypt & N. Africa \\ \bottomrule \end{tabular}
            \end{adjustbox} \\ \hline
            \begin{adjustbox}{width=\linewidth}
            \begin{tabular}{@{}cccccc@{}} \toprule Item & Discovery & Artifact & Date & Culture & Region \\ \midrule Scroll & Giza Pyramid & Diamond & \textasciitilde2500 BC & Anc. Egypt & N. Africa \\ Knife & Tut. Tomb & Stone Tab. & 1323 BC & Anc. Egypt & N. Africa \\ \bottomrule \end{tabular}
            \end{adjustbox} \\ \hline
            \end{tabular}
        \end{adjustbox}
        \caption{UGEN$_{\text{V1}}$ Example: Two archaeology tables with identical information and permuted but perfectly alignable columns, incorrectly labeled non-unionable despite clear semantic compatibility.}
        \label{tab:ugen_incomplete_v1_app}
    \end{subfigure}
    \begin{subfigure}{1.0\columnwidth}
        \centering
        \begin{adjustbox}{width=\columnwidth,center}
            \small
            \renewcommand{\arraystretch}{1.1} \setlength{\tabcolsep}{3pt}
            \begin{tabular}{|p{\linewidth}|} \hline
            \makecell[l]{\textbf{Query:} Veterinary-Science\_YP1NJGLN.csv \\ \textbf{Candidate:} Veterinary-Medicine\_GVNM098Q.csv} \\ \hline
            \begin{adjustbox}{width=\linewidth}
            \setlength{\tabcolsep}{2.5pt}
            \begin{tabular}{@{}llllll@{}} \toprule Animal Type & Breed & Age & Health Status & Symptoms & Diagnosis \\ \midrule Dog & Labrador Retr. & 3 years & Healthy & No symptoms & Routine check-up \\ Cat & Domestic SH & 5 years & Overweight & Lethargy... & Obesity \\ \bottomrule \end{tabular}
            \end{adjustbox} \\ \hline
            \begin{adjustbox}{width=\linewidth}
            \setlength{\tabcolsep}{2.5pt}
            \begin{tabular}{@{}llllll@{}} \toprule Animal Type & Breed & Age & Gender & Symptoms & Diagnosis \\ \midrule Dog & Labrador & 3 years & Male & Aggression... & Rabies \\ Cat & Siamese & 8 years & Female & Limping... & Arthritis \\ \bottomrule \end{tabular}
            \end{adjustbox} \\ \hline
            \end{tabular}
        \end{adjustbox}
        \caption{UGEN$_{\text{V2}}$ Example: Two veterinary case tables with highly alignable core columns (Animal Type, Breed, Age, Symptoms, Diagnosis) representing the same fundamental entity type (animal patients).}
        \label{tab:ugen_incomplete_v2_app}
    \end{subfigure}
    \caption{Examples of \textsc{Ugen} Pairs explicitly labeled as non-unionable in the original ground truth exhibiting strong compatibility suggesting unionability.}
    \label{fig:ugen_incorrect_negatives_app}
\end{figure}

\subsection{Lakebench Benchmark Inconsistencies}\label{appendix:lb_examples}

This subsection presents examples of GTFPs from the \textsc{LakeBench} benchmarks, where semantically and structurally compatible tables were not labeled as unionable in the ground truth but were correctly retrieved by search methods. Figures \ref{fig:lakebench_webtable_incorrect_negatives_app} and \ref{fig:lakebench_opendata_incorrect_negatives_app} show such cases from the WebTable and OpenData subsets, respectively.
\begin{figure}[htbp] 
    \centering
    \begin{subfigure}{1.0\columnwidth}
        \centering
        \begin{adjustbox}{width=\columnwidth,center}
            \small
            \renewcommand{\arraystretch}{1.1} \setlength{\tabcolsep}{3pt}
            \begin{tabular}{|p{\linewidth}|} \hline
            \makecell[l]{\textbf{Query:} csvData10212811.csv \\ \textbf{Candidate:} csvData1066748.csv} \\ \hline
            \begin{adjustbox}{width=\linewidth}
             \setlength{\tabcolsep}{2.5pt} 
            \begin{tabular}{@{}llccccccc@{}} \toprule Player & Team & POS & G & AB & H & HR & ... & OPS \\ \midrule 
            B Dean & GL & 1B & 96 & 350 & 83 & 7 & ... & 0.657 \\ 
            Y Arbelo & SB & 1B & 134 & 461 & 114 & 31 & ... & 0.877 \\ 
            \bottomrule \end{tabular}
            \end{adjustbox} \\ \hline
            \begin{adjustbox}{width=\linewidth}
             \setlength{\tabcolsep}{2.5pt} 
            \begin{tabular}{@{}llccccccc@{}} \toprule Player & Team & POS & G & AB & H & HR & ... & OPS \\ \midrule 
            J Colina & WS & 2B & 59 & 216 & 66 & 3 & ... & 0.832 \\ 
            B Friday & LYN & SS & 85 & 341 & 98 & 2 & ... & 0.752 \\ 
            \bottomrule \end{tabular}
            \end{adjustbox} \\ \hline
            \end{tabular}
        \end{adjustbox}
        \caption{WebTable Example 1: Baseball player statistics tables with identical, rich schemas (including Player, Team, POS, G, AB, H, HR, OPS, etc.). These tables represent the same entity type (player season stats) and are highly unionable, but were not labeled as such in the ground truth.}
        \label{tab:lakebench_webtable_fn1_app_new} 
    \end{subfigure}
    \begin{subfigure}{1.0\columnwidth}
        \centering
        \begin{adjustbox}{width=\columnwidth,center}
            \small
            \renewcommand{\arraystretch}{1.1} \setlength{\tabcolsep}{3pt}
            \begin{tabular}{|p{\linewidth}|} \hline
            \makecell[l]{\textbf{Query:} csvData10025189.csv \\ \textbf{Candidate:} csvData20099586.csv} \\ \hline
            \begin{adjustbox}{width=\linewidth}
             \setlength{\tabcolsep}{2.5pt} 
            \begin{tabular}{@{}lcccccccc@{}} \toprule Player & Team & POS & AVG & G & AB & R & ... & OPS \\ \midrule 
            A Ramirez & MIL & 3B & 0.285 & 133 & 494 & 47 & ... & 0.757 \\ 
            E Chavez & ARI & 3B & 0.246 & 44 & 69 & 6 & ... & 0.795 \\ 
            \bottomrule \end{tabular}
            \end{adjustbox} \\ \hline
            \begin{adjustbox}{width=\linewidth}
             \setlength{\tabcolsep}{2.5pt} 
            \begin{tabular}{@{}lcccccccc@{}} \toprule Player & Team & POS & AVG & G & AB & R & ... & OPS \\ \midrule 
            L Castillo & NYM & 2B & 0.245 & 87 & 298 & 46 & ... & 0.660 \\ 
            R Durham & MIL & 2B & 0.289 & 128 & 370 & 64 & ... & 0.813 \\ 
            \bottomrule \end{tabular}
            \end{adjustbox} \\ \hline
            \end{tabular}
        \end{adjustbox}
        \caption{WebTable Example 2: More baseball player statistics tables with identical schemas, clearly unionable but not labeled as such.}
        \label{tab:lakebench_webtable_fn2_app}
    \end{subfigure}
    \caption{Examples of \textsc{LB-Webtable} Ground Truth Incompleteness.}
    \label{fig:lakebench_webtable_incorrect_negatives_app}
\end{figure}

\begin{figure}[t] 
    \centering
    \begin{subfigure}{1.0\columnwidth}
        \centering
        \begin{adjustbox}{width=\columnwidth,center}
            \small
            \renewcommand{\arraystretch}{1.1} \setlength{\tabcolsep}{3pt}
            \begin{tabular}{|p{\linewidth}|} \hline
            \makecell[l]{\textbf{Source:} OpenData (Canada) \\ \textbf{Query:} CAN\_CSV0000000000000659.csv \\ \textbf{Candidate:} CAN\_CSV0000000000000562.csv} \\ \hline
            \begin{adjustbox}{width=\linewidth}
             \setlength{\tabcolsep}{2.5pt} 
            \begin{tabular}{@{}lp{0.25\linewidth}llc@{}}
            \toprule REF\_DATE & GEO & Age group & Sex & ... VALUE \\
            \midrule 2003 & Canada & Total, 12 years and over & Both sexes & ... 20723896.0 \\
            2003 & Canada & Total, 12 years and over & Both sexes & ... 20632799.0 \\
            \bottomrule
            \end{tabular}
            \end{adjustbox} \\ \hline
            \begin{adjustbox}{width=\linewidth}
             \setlength{\tabcolsep}{2.5pt} 
            \begin{tabular}{@{}lp{0.25\linewidth}llc@{}}
            \toprule REF\_DATE & GEO & Age group & Sex & ... VALUE \\
            \midrule 2003 & Canada & Total, 12 years and over & Both sexes & ... 26567928.0 \\
            2003 & Canada & Total, 12 years and over & Both sexes & ... 26567928.0 \\
            \bottomrule
            \end{tabular}
            \end{adjustbox} \\ \hline
            \end{tabular}
        \end{adjustbox}
        \caption{OpenData Example 1: Canadian health survey tables sharing key demographic columns (REF\_DATE, GEO, Age group, Sex) for the same population. This pair represents unionable statistics about that population but was not labeled as unionable in the ground truth.}
        \label{tab:lakebench_opendata_fn1_app} 
    \end{subfigure}
    \begin{subfigure}{1.0\columnwidth}
        \centering
        \begin{adjustbox}{width=\columnwidth,center}
            \small
            \renewcommand{\arraystretch}{1.1} \setlength{\tabcolsep}{3pt}
            \begin{tabular}{|p{\linewidth}|} \hline
            \makecell[l]{\textbf{Source:} OpenData (Canada) \\ \textbf{Query:} CAN\_CSV0000000000000686.csv \\ \textbf{Candidate:} CAN\_CSV0000000000005304.csv} \\ \hline
            \begin{adjustbox}{width=\linewidth}
             \setlength{\tabcolsep}{2pt} 
            \begin{tabular}{@{}lp{0.26\linewidth}p{0.29\linewidth}llc@{}} 
            \toprule Sex & Type of work & Hourly wages & UOM & UOM\_ID & ... VALUE \\
            \midrule 
            Both & Both full- and part... & Total employees, all wages & Persons & 249 & ... 10921.0 \\
            Males & Both full- and part... & Total employees, all wages & Persons & 249 & ... 5645.4 \\
            \bottomrule
            \end{tabular}
            \end{adjustbox} \\ \hline
            \begin{adjustbox}{width=\linewidth}
             \setlength{\tabcolsep}{2pt} 
            \begin{tabular}{@{}lp{0.26\linewidth}p{0.29\linewidth}llc@{}} 
            \toprule Sex & Type of work & Weekly wages & UOM & UOM\_ID & ... VALUE \\
            \midrule 
            Both & Both full- and part... & Total employees, all wages & Persons & 249 & ... 11364.5 \\
            Males & Both full- and part... & Total employees, all wages & Persons & 249 & ... 5954.5 \\
            \bottomrule
            \end{tabular}
            \end{adjustbox} \\ \hline
            \end{tabular}
        \end{adjustbox}
        \caption{OpenData Example 2: Canadian employment statistics. The query table (data related to 'Hourly wages') and candidate table (data related to 'Weekly wages') share key dimensions like Sex, Type of work, and UOM. The cell values within their respective 'Hourly wages'/'Weekly wages' columns (e.g., 'Total employees, all wages') describe similar employee groups. This pair, differing mainly in wage aggregation period (hourly vs. weekly) and slightly in REF\_DATE format (YYYY vs YYYY-MM), is potentially unionable for comprehensive wage analysis but was not labeled as such in the ground truth.}
        \label{tab:lakebench_opendata_fn2_app_new} 
    \end{subfigure}
    \caption{Examples of \textsc{LB-OpenData} Ground Truth Incompleteness.}
    \label{fig:lakebench_opendata_incorrect_negatives_app}
\end{figure}

\section{LLM Adjudicator} \label{appendix:prompt}

\subsection{Prompt Details}
To systematically re-evaluate potential ground truth inconsistencies in the \textsc{Ugen} benchmarks, we employed an LLM-based adjudicator. This process targeted disagreements identified during our analysis, specifically Ground Truth False Positives (GTFPs, pairs retrieved as potentially unionable within a rank threshold $k'$ but not labeled as unionable in the ground truth, $k' < k$) and Ground Truth False Negatives (GTFNs, pairs labeled as unionable in the ground truth but retrieved within a rank threshold $k'$, $k' > k$, or not retrieved at all).

For each query-candidate pair under review, we provided the LLM with the full content of both tables. The table data was serialized into a Markdown format using the \texttt{MarkdownRawTableSerializer} recipe from the Table Serialization Kitchen library\footnote{\href{https://github.com/daniel-gomm/table-serialization-kitchen}{Table Serialization Kitchen Github Repository}}\citep{gomm2025metadata}. This serialized data was inserted into specific placeholders \texttt{(`<Query Table Data>`, `<Candidate Table Data>`)} within the prompt detailed below. Crucially, the original table names were \textit{not} included in the prompt. This decision was made to avoid potentially biasing the LLM by providing explicit hints about the table's topic beforehand, thereby ensuring the adjudication relies solely on the semantic and structural information present in the table content itself.

The prompt utilizes few-shot learning, incorporating hand-selected positive and negative examples of unionability from the \textsc{Ugen} benchmarks themselves to guide the LLM's judgment (these examples are represented by a placeholder in the verbatim prompt below for brevity). The prompt defines the LLM's role, outlines core principles for assessing conceptual coherence and semantic column alignment, and specifies the required output format.

The complete prompt structure provided to the LLM adjudicator is shown below:

\begin{lstlisting}
You are an experienced data curator evaluating if two database tables can be meaningfully combined vertically (unioned). The goal of unioning is to create a single, larger dataset containing the same kind of information or describing the same type of entity/event.

Your task is to determine if TABLE 1 and TABLE 2 are conceptually compatible enough for a union operation.

CORE PRINCIPLES FOR UNIONABILITY:

1.  Conceptual Coherence: Do both tables fundamentally describe the same type of entity (e.g., customers, products, logs) or record the same type of event (e.g., sales, website visits)? Appending rows from one table to the other should result in a dataset that makes logical sense.

2.  Meaningful Column Alignment: There must be a reasonable set of columns across the two tables that represent the same underlying attributes or concepts.
    * These columns can have DIFFERENT NAMES (e.g., "Cust_ID" vs. "ClientIdentifier").
    * They can have DIFFERENT FORMATS (e.g., "2023-01-15" vs. "1/15/2023").
    * They may have LITTLE TO NO OVERLAP in actual data values.
    * Focus on the semantic meaning of the columns in the context of their respective tables.

3. Sufficient Column Matching: The alignment shouldn't rely on just one incidental or minor column. There should be enough matching among key columns to confidently conclude that the tables represent the same underlying information. More aligned columns representing core attributes increase confidence.

4.  Distinction from Joins: We are NOT looking for keys to join tables horizontally. We are assessing if they can be stacked vertically.

[EXAMPLES OF UNIONABILITY/NON-UNIONABILTY:]
(Hand-selected positive/negative examples from \textsc{Ugen} benchmark were inserted here in the actual prompt)

YOUR TASK:

Examine the following two tables based on the principles and examples above:

TABLE 1:
<Query Table Data (Serialized)>

TABLE 2:
<Candidate Table Data (Serialized)>

PROVIDE YOUR ANSWER IN THE FOLLOWING FORMAT:

1.  First line: `UNIONABLE: Yes` or `UNIONABLE: No`
2.  Second line: `EXPLANATION:` followed by a brief justification focusing on the conceptual coherence and the sufficiency of semantic column alignment. Explain why they are or are not the same kind of data.
\end{lstlisting}

\subsection{Example LLM Outputs}

Below are illustrative examples of the outputs generated by the LLM adjudicator based on the prompt above for some table pairs showcased in Appendix ~\ref{appendix:ugen_examples}.

\paragraph{Example 1:} Archeology\_2LWSQ5A2.csv vs. Archeology\_3ML53C0M.csv (UGEN-V1)
\begin{lstlisting}
UNIONABLE: Yes
EXPLANATION: Both tables describe archeological artifacts and their attributes (discovery location, item type, date, culture, material, etc.). They represent the same kind of data - records of archeological findings. Unioning them would create a consolidated dataset of such findings. There is strong semantic alignment across almost all columns, even though the column order differs.
\end{lstlisting}

\paragraph{Example 2:} Anthropology\_FGTNBDWF.csv vs. Anthropology\_N30U114M.csv (UGEN-V1)
\begin{lstlisting}
UNIONABLE: No
EXPLANATION: TABLE 1 categorizes anthropological concepts based on dimensions like 'Age', 'Culture', 'Arena', 'Domain', 'Meaning', 'Origin', and 'Activity'. TABLE 2, on the other hand, lists categories such as 'Artifact', 'Language', 'Technology', 'Education', and 'Society' as columns. While both tables are related to Anthropology, they represent different perspectives and categorizations within the field. There is insufficient semantic overlap in the columns to suggest they describe the same type of entity or event. Unioning them would likely result in a table that is not conceptually coherent, as the rows from each table represent different kinds of anthropological classifications.
\end{lstlisting}

\end{document}